\documentclass[twocolumn]{aastex631}

\usepackage{CJK}

\shorttitle{Variable Accretion onto the nearby young star TW Hya}
\shortauthors{Ji et al.}

\begin{document}
\begin{CJK*}{UTF8}{gbsn}
\title{The mass distribution of clumpy accretion onto the nearby young star TW Hya}


\author[0009-0004-2113-3096]{Tao JI (纪涛)*}
\affiliation{Department of Physics, The University of Hong Kong, Pokfulam Road, Hong Kong}
\affiliation{Department of Astronomy, Peking University, Yiheyuan Lu 5, Haidian Qu, 100871 Beijing, People's Republic of China}

\author[0000-0001-7351-6540]{Javier Serna}
\affiliation{Homer L. Dodge Department of Physics and Astronomy, University of Oklahoma, Norman, OK 73019, USA}

\author[0000-0002-7154-6065]{Gregory J. Herczeg (沈雷歌）*}
\affiliation{Department of Astronomy, Peking University, Yiheyuan Lu 5, Haidian Qu, 100871 Beijing, People's Republic of China}
\affiliation{Kavli Institute for Astronomy and Astrophysics, Peking University, Yiheyuan Lu 5, Haidian Qu, 100871 Beijing, People's Republic of China}

\author[0000-0003-3882-3945]{Shinsuke Takasao}
\affiliation{Humanities and Sciences/Museum Careers, Musashino Art University, Tokyo 187-8505, Japan}

\author[0000-0001-7796-1756]{Frederick M. Walter}
\affiliation{Department of Physics and Astronomy, Stony Brook University, StonyBrook, NY 11794, USA}

\author[0000-0003-4520-5395]{Yuguang Chen (陈昱光)}
\affiliation{Department of Physics, The Chinese University of Hong Kong, Shatin, N.T., Hong Kong}

\author[0000-0002-8533-168X]{Antonio Armeni}
\affiliation{Institut f\"ur Astronomie und Astrophysik, Eberhard Karls Universität Tübingen, Sand 1, 72076 Tübingen, Germany}

\author[0000-0002-6773-459X]{Doug Johnstone}
\affiliation{NRC Herzberg Astronomy and Astrophysics, 5071 West Saanich Road, Victoria, BC V9E 2E7, Canada}
\affiliation{Department of Physics and Astronomy, University of Victoria, Victoria, BC V8W 2Y2, Canada}

\author[0000-0001-6496-0252]{Jochen Eisl\"offel}
\affiliation{Th\"uringer  Landessternwarte Tautenburg, Sternwarte 5, D-07778 Tautenburg, Germany}

\author[0000-0001-8060-1321]{Min Fang}
\affiliation{Purple Mountain Observatory, Chinese Academy of Sciences, 10 Yuanhua Road, Nanjing 210023, People's Republic of China}
\affiliation{School of Astronomy and Space Science, University of Science and Technology of China, Hefei 230026, People's Republic of China}

\author[0000-0001-9590-2274]{Sean P. Matt}
\affiliation{Homer L. Dodge Department of Physics and Astronomy, University of Oklahoma, Norman, OK 73019, USA}

\author[0000-0001-5018-3560]{Michal Siwak}
\affiliation{Mt. Suhora Astronomical Observatory, University of the National Education Commission, ul. Podchorkazych 2, 30-084 Krakow, Poland}

\author[0000-0002-4115-0318]{Laura Venuti}
\affiliation{SETI Institute, 339 Bernardo Ave., Suite 200, Mountain View, CA 94043, USA}

\author[0000-0002-4147-3846]{Miguel Vioque}
\affiliation{European Southern Observatory, Karl-Schwarzschild-Str. 2, 85748 Garching bei M\"unchen, Germany}

\author[0000-0002-9589-5235]{Lixin Dai}
\affiliation{Department of Physics, The University of Hong Kong, Pokfulam Road, Hong Kong}
\footnote{Corresponding authors: Gregory Herczeg, gherczeg1@gmail.com, and Tao Ji, jitao@pku.edu.cn}

\begin{abstract}
The proliferation of high time-resolution and decades-long monitoring of classical T Tauri stars provides a vast opportunity to test the variability of the star-disk connections.  However, most monitoring surveys use single broad-band filters,  which makes the conversion of photometric variability into accretion rate difficult.  In this study, we analyze accretion bursts onto the nearby young star TW Hya over short (hours, days) and long (months, years) timescales by calibrating TESS and ASAS-SN $g$-band photometry to accretion rates with simultaneous spectroscopy.  The high cadence TESS light curve shows bursts of accretion in clumps with masses from a sensitivity limit of $\sim10^{-13}$~M$_\odot$ up to
$3\times 10^{-11}$\,M$_\odot$.
The average burst duration of 1.8 days is longer than a simple estimate of the thermal response timescale, supporting the interpretation that the photometric variability probes the instantaneous accretion rate.
The reset timescale of 1.2--2 days derived from the structure function and previously reported quasi-periods of 3.5--4 days are consistent with bursts that may be related to the different rotation between the stellar magnetosphere and inner disk or with azimuthal asymmetries in the inner disk.  The near-daily ASAS-SN light curve across 8 years reveals some seasonal changes in brightness with a standard deviation of $\sim 0.13$ mag, about half of the scatter seen on short timescales.  This study demonstrates the importance of coordinating contemporaneous multi-epoch spectroscopy with time domain surveys to interpret light curves of young stars.
\end{abstract}


\section{Introduction}

In the magnetospheric accretion paradigm for classical T Tauri stars, gas from the disk accretes onto the star along magnetic field lines \citep{hartmann16}.  This star-disk connection is inherently unstable \citep[e.g.][]{romanova08}.  On timescales of years to decades, a build-up of material near the magnetospheric boundary or the dead zone inner boundary may trigger abrupt accretion outbursts \citep{dangelo10,lee20,cecil24}.  On hours-to-days timescales, localized instabilities control the amount of gas that reaches the footpoint of the star-disk connection \citep{blinova16,takasao22,zhu24}, possibly inducing accretion along multiple streams \citep{romanova25}.

Accretion variability on the hours-to-days timescales have been best revealed with space-based  high cadence photometric monitoring, beginning with COROT and MOST and continuing with K2 and now TESS \citep[e.g.][]{rucinski08,alencar10,siwak11,stauffer14,cody18,robinson22}. From these datasets, accretion variability onto young stars is usually characterized as bursty, sometimes including quasi-periodic signals.  For some stars the accretion variability is mixed with changes in extinction through the disk and visible spot coverage (e.g., \citealt{venuti21} and \citealt{cody22}, following the framework developed by \citealt{herbst94}).
The associated timescales are consistent with variability seen in simulations of magnetospheric accretion \citep{romanova25}.  However, the mass associated with these short-timescale bursts is usually not quantified because converting photometric changes into accretion rate variability is complicated, especially without ancillary supporting data.

In this paper, we reassess the photometric variability of TW Hya, with a focus on quantifying the mass accreted during short bursts and searching for any long-term trends across months and years. Studies of TW Hya, one of the closest protoplanetary disk systems (60 pc, \citealt{gaiadr3}), have helped to form a foundation for our understanding of magnetospheric accretion. The star and disk are seen nearly pole-on \citep[e.g.][]{qi05,donati11}, so the optical brightness should not be affected by extinction events and is minimally affected by spot rotation.  Time-series analysis of high-resolution photometry shows frequent bursts, with characteristic timescales of $\sim 4$ days \citep[e.g.][]{rucinski08,siwak14,siwak18}.  Spectroscopy and photometry obtained over the past two decades indicates that TW Hya has relatively stable accretion, with a reset timescale of $\sim 1.6$ days, a slowly varying accretion footprint and variability on days-long timescales \citep{herczeg23,sicilia23,donati24,romanova25}.

The high cadence optical light curve of TW Hya is punctuated by frequent short bursts \citep[e.g.][]{sicilia23,wendeborn24}, as is commonly seen from most accreting young stars \citep[e.g.][]{cody22}.  Any modest variability in the long-term light curve may be masked by the scatter introduced by the bursts.
In this paper, we provide the first measurements of the mass accreted in these bursts and evaluate the long-term variability of this accreting system.  In Section \ref{section:data}, we introduce the photometry and spectroscopy datasets used in this paper.  In Section \ref{section:correlation}, we correlate photometric variability with mass accretion rates.  In Section \ref{section:results}, we use those correlations to measure the properties of bursts and assess long-term changes in accretion.
In Section \ref{section:discussion}, we interpret this variability and discuss how this framework may be applied to other stars.
Finally, we summarize the results in Section \ref{sec:summary}.

\section{Data}\label{section:data}
This section presents the datasets analyzed, providing a brief overview of necessary data reductions. Detailed descriptions are provided for TESS photometry (Section \ref{section:data_tess}), ASAS-SN photometry (Section \ref{section:data_asassn}), synthetic photometry (Section \ref{section:synthetic}), and high-resolution spectroscopy (Section \ref{section:data_echelle}).

\subsection{TESS}\label{section:data_tess}

\begin{figure*}[!t]
\centering
\includegraphics[scale=0.21]
{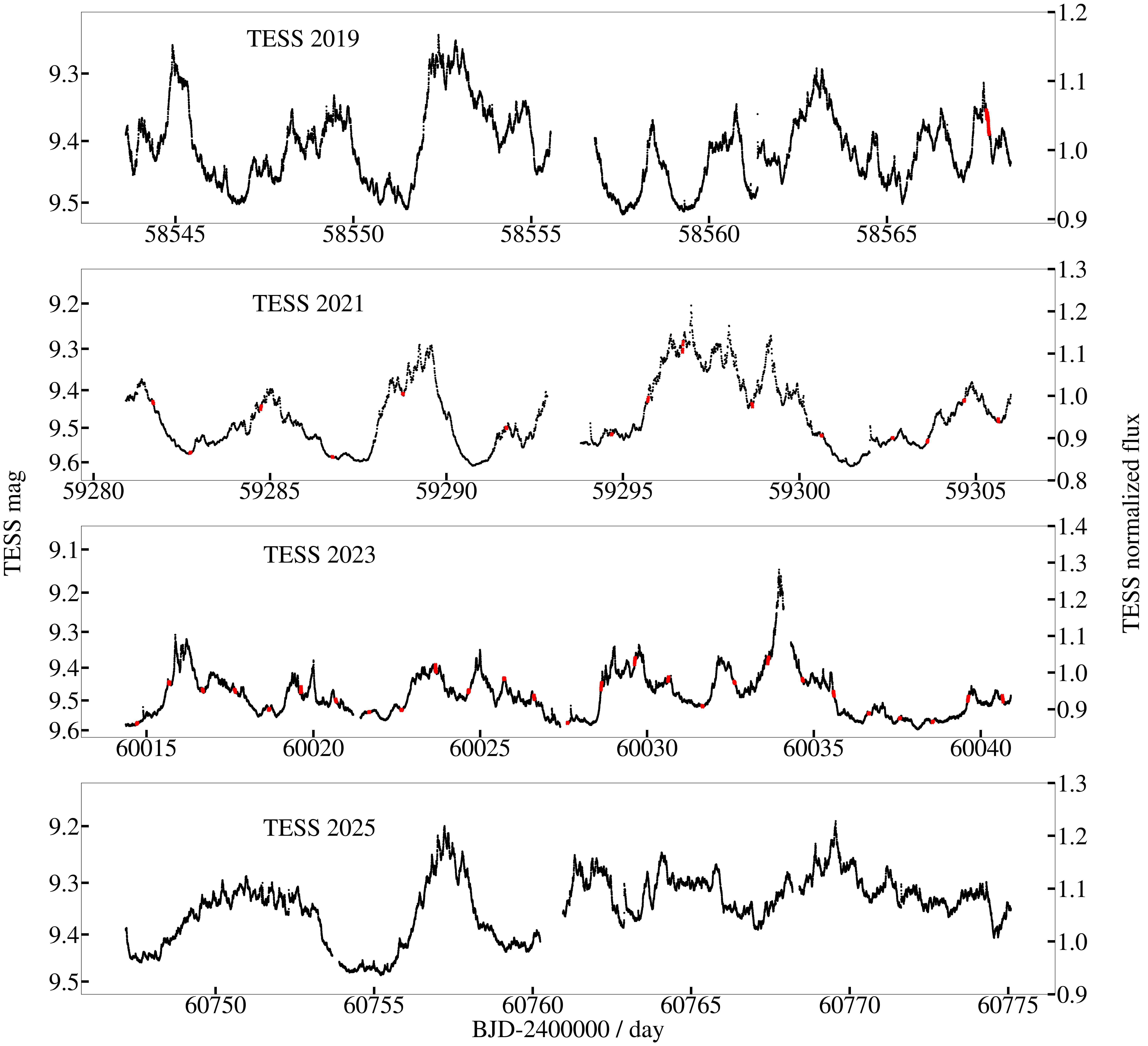}
\caption{The 25-day TESS light curves of TW Hya from (top to bottom) 2019, 2021, 2023, and 2025, showing frequent bursts.  Red crosses mark the data points with simultaneous accretion rates from high-resolution spectroscopy (see Section \ref{section:correlation_tess}).  The TESS flux, normalized by the median ($22890 \ {\rm e^{-} / s}$, $9.41$ mag) of these four lightcurves is plotted on the right y-axis.}
\label{fig:tess}
\end{figure*}

The TESS (Transiting Exoplanet Survey Satellite) mission provides high-quality photometry for bright targets, with the primary goal of discovering transiting exoplanets \citep{ricker15}. TESS observes most regions for $\sim25$ day windows once every two years.  
 TW Hya was observed with TESS Camera 2 in 2019 from March 1 to March 25, 2021 from March 7 to April 1, 2023 from March 10 to April 6 and 2025 from March 12 to April 9. The monitoring cadence of TW Hya is 2 minutes for the 2019, 2023 and 2025 observations and 10 minutes for the 2021 observations (see Figure \ref{fig:tess}).  The TESS data are analyzed using normalized fluxes, with flux units normalized by the median of 22890~$e^-/s$ in the four light curves.

The filter transmission curve of TESS covers $600-1040$ nm. The photometry data was reduced using the TESS SPOC pipeline \citep{caldwell20}. No other bright sources are located within the large TESS 21$^{\prime\prime}$ pixels, so the light curve should not be contaminated.

\subsection{ASAS-SN}\label{section:data_asassn}

\begin{figure*}[t]
\centering
\includegraphics[scale=0.9]
{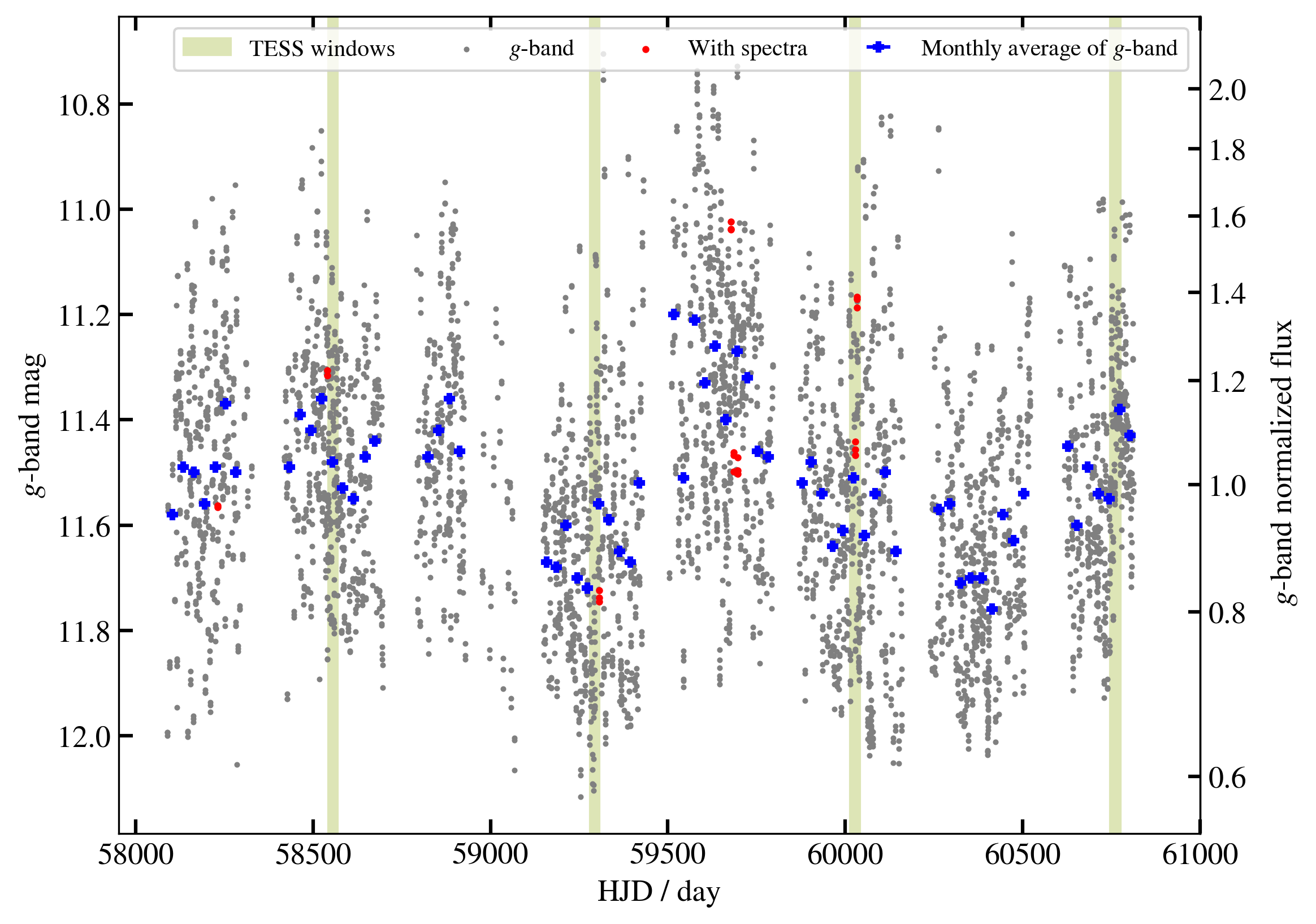}
\caption{The 2700-day $g$-band photometry of TW Hya from ASAS-SN, with approximately daily observations from 2017 Dec to 2025 May (gray points; right y-axis shows flux normalized by the median brightness of 11.52 mag), with simultaneous accretion rate measurements from high-resolution spectroscopy (red points) and shaded regions marking the four 25-day TESS observing windows.
Short-term fluctuations on day timescales are larger than the variability of monthly averages (blue points).}
\label{fig:asassn}
\end{figure*}

The All-Sky Automated Survey for Supernovae (ASAS-SN) is a long-term program monitoring the entire visible sky almost every night \citep{shappee14, kochanek17}. The long-term ASAS-SN light curves complement the short-cadence TESS light curves.  The ASAS-SN network currently consists of 24 telescopes distributed across the globe, with sensitivity to $\sim 18$ mag for $g$-band and $\sim 17$ mag for $V$-band. 

We use only the $g$-band observations, with 4754 distinct $g$-band photometric points spread over $\sim 2700$ days, from 2017 Dec -- 2025 May. The ASAS-SN light curves are analyzed in linear space (flux rather than magnitudes), normalized by the median photometric magnitude of 11.52.  The $g$-band observations were obtained using three dithered $90$\,s exposures to increase the signal-to-noise ratio of the light curves.  Zero points are calibrated from the AAVSO Photometric All-Sky Survey (APASS) catalogue \citep[see reduction and calibration details in][]{shappee14, kochanek17}.

\subsection{Synthetic photometry from spectroscopy}\label{section:synthetic}

\citet{herczeg23} measured accretion rates for TW Hya from 26 low-resolution, flux-calibrated spectra by fitting the excess hydrogen continuum emission, including the Balmer Jump. In order to increase the number of datapoints with $g$-band photometry and spectroscopic accretion rates,  we calculate synthetic $g$-band photometry for the 18 spectra with the most reliable flux calibration, obtained from HST/STIS, Keck/LRIS, Keck/HIRES (high resolution), and four VLT/X-Shooter spectra.  The $g$-band filter transmission curve is obtained from the Spanish Virtual Observatory\footnote{http://svo.laeff.esa.es}.  The synthetic $g$-band photometry is shown in Table \ref{tab:synthetic} in the  Appendix \ref{appendix:synthetic}.

\subsection{High resolution spectroscopy}\label{section:data_echelle}

We convert the TESS and ASAS-SN $g$-band photometry into a time series of accretion rate by establishing a relationship between the photometry and simultaneous accretion rates from high-resolution spectroscopy.  
The accretion rates were obtained by measuring the veiling in 1167 high-resolution spectra, with a full description in \cite{herczeg23}.  We supplement those accretion rates with 31 additional CHIRON spectra obtained from 2023 January 16 through 2023 April 7 (PI Walter), including 27 that were obtained to overlap with the TESS monitoring campaign (see Appendix \ref{appendix:veiling}).
  Many of these spectra were also coordinated with UV spectra obtained in the ULLYSES program \citep{ullyses25}, as part of ODYSSEUS collaboration \citep{espaillat22}.

The veiling was converted into accretion rates using hydrogen slab models developed by \citet{valenti93}, assuming no change in the underlying photospheric emission.  The random error in veiling measurements from 5000--5100 \AA\ is about 0.01, following the analysis of CHIRON spectra by \citet{herczeg23}.

\begin{figure*}[!t]
\centering
\includegraphics[scale=0.52]
{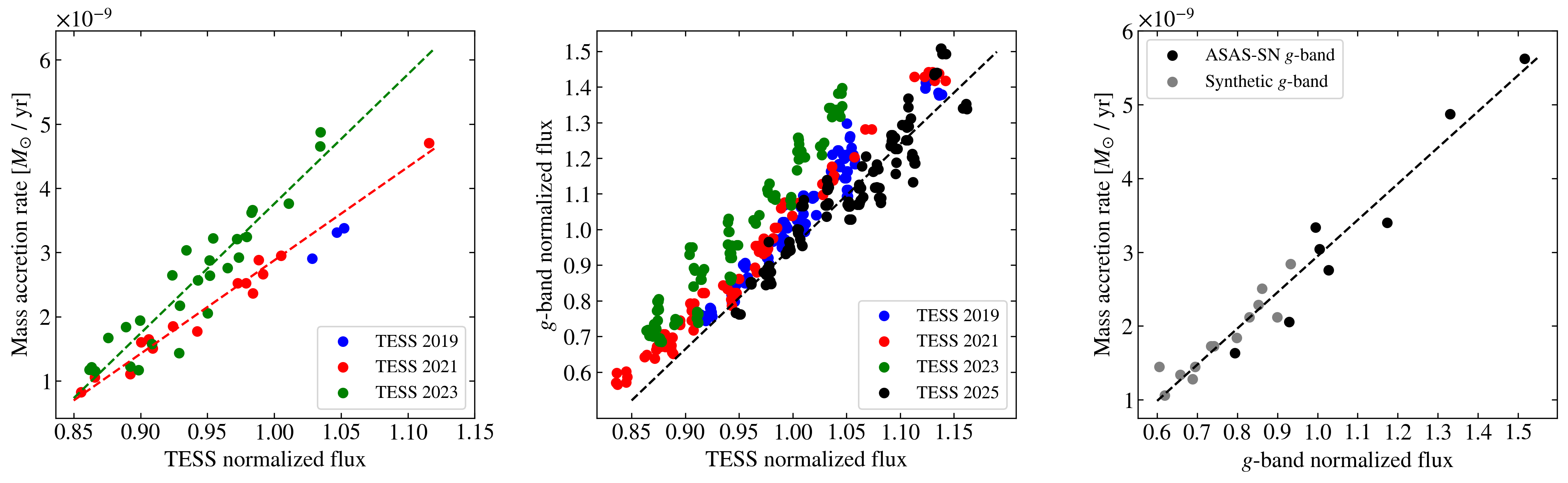}
\caption{Linear relationships (dashed lines) between TESS photometry and simultaneous accretion rate measurements from high resolution spectroscopy (left), the g-band and TESS fluxes (middle), and accretion rate and g-band flux (right).  The different epochs of TESS are identified with different colors and are fit separately.}
\label{fig:linear_combined}
\end{figure*}

\section{Establishing relationships between photometry and accretion rate}\label{section:correlation}

Figures \ref{fig:tess}-\ref{fig:asassn} show the TESS and ASAS-SN $g$-band light curves.  The TESS monitoring reveals frequent bursts.  The ASAS-SN data appears noisy because those bursts are sampled about once per day.  The emission consists of a cool photosphere and blue accretion \citep[e.g.][]{manara14}, so bluer wavelengths (in this paper, ASAS-SN $g$-band) are more sensitive than red wavelengths (in this paper, TESS) to changes in accretion.  For TW Hya, \citet{herczeg23} found that veiling can increase by 1.5 (a change of 1 mag) at 4450 \AA\ but only $\sim 0.4$ at 7050 \AA\ (0.36 mag), roughly consistent with the full range of $g$-band and TESS magnitudes.

In this section, we first find strong linear correlations to convert these photometric light curves to accretion rate curves and estimate the photospheric contribution to brightness.
The tight correlations confirm expectations that the TESS and ASAS-SN $g$-band photometric variability is dominated by accretion variability, consistent with previous analyses \citep[e.g.][]{wendeborn24}.

In these correlations, the normalized fluxes corresponding to $9.41$ mag for TESS and $11.52$ mag for $g$-band are
used  because the relationships are in linear rather than magnitude space.
The TESS photometry is averaged in 20-minute windows around the time of the spectroscopic observation to cover the typical spectroscopic integration time.  

The accretion rates presented here could be systematically underestimated by a factor of $\sim 1.5$.   When modeling the spectrum of TW Hya, \citet{espaillat19} used multiple accretion flows with a range of densities, leading to higher accretion luminosities than those obtained from hydrogen slab models (the method implicit in our analysis, based on \citealt{herczeg23}, and also adopted in X-shooter analysis, e.g., \citealt{manara14}).
This systematic uncertainty is not considered further but would be relevant to any quantified comparison to simulations.

\vspace{10mm}

\subsection{Correlations between TESS brightness and accretion rate}\label{section:correlation_tess}

Strong linear correlations between TESS brightness and accretion rate (Figure~\ref{fig:linear_combined} and Table~\ref{table:fitting}) allow us to convert the photometric light curves into a time series of accretion rate.
The spectroscopic measurements in the 2021 and 2023 visits trace all phases in the photometric states of TW Hya, including pre-bursts, during bursts, and post-burst epochs (see red crosses in Figure \ref{fig:tess}).   The scatter in linear regressions (Table \ref{table:fitting}) is 7--15\%, with separate fits for 2021 (15 points of overlap between TESS and accretion rate measurements) and 2023 (27 points of overlap).  For the 2019 TESS photometry, we adopt the linear fit from the 2021 data based on consistency with the three simultaneous TESS and spectroscopy data points.

All four TESS epochs also show strong linear relationships between photometry and ASAS-SN $g$-band brightness (see Table \ref{table:fitting} and middle panel in Figure \ref{fig:linear_combined}).  
 The primary difference in the fits is the y-intercepts.  Modest differences in the slopes lead to relatively small differences in the maximum minus minimum values measured in the figure.  These relationships confirm the similarity between the 2019 and 2021 TESS conversions.  For 2025, the relationship between TESS magnitude and $g$-band is used to calculate mass accretion rates, given the lack of spectroscopic coverage.  The accretion rates estimated for the 2025 TESS observations are higher than in other years, as inferred from the brighter $g$-band photometry.

For TW Hya, the face-on geometry means that extinction does not vary and changes in spot coverage are minor.   The 2021 and 2023 slopes differ due to changes either in the visible spot coverage or in the temperature and density distribution of the accretion flow \citep[see also discussions in][]{robinson22,wendeborn24}.  Such differences are expected empirically from the higher scatter seen in correlations between veiling at red and blue wavelengths from optical spectra, as described in \cite{herczeg23}.    These correlations are adopted for TW Hya and should not be applied to other sources.

\subsection{Correlations between ASAS-SN g-band brightness and accretion rate}\label{section:correlation_gband}

To address the long-term evolution of accretion rate and to scale the 2025 TESS photometry to accretion rate,
we correlate the daily ASAS-SN g-band photometry data with mass accretion rate measured from veiling (right panel in Figure \ref{fig:linear_combined}).  These data come from the $22$ times ASAS-SN $g$-band photometry obtained within $20$ minutes of spectroscopy. We supplement this empirical correlation by adding synthetic $g$-band photometry data with accretion rates from the low-resolution spectra from \cite{herczeg23} with the most accurate flux calibrations. 

The correlation between $g$-band brightness and accretion rate has an rms scatter of 11\% in accretion rate.  After subtracting off the photospheric level (see \S 3.3), doubling the normalized flux leads to doubling the accretion rate.  The relationship spans from $1-5.7\times10^{-9}$ M$_\odot$ yr$^{-1}$.  At lower accretion rates, the uncertainty in photospheric emission (here the x-intercept of 0.4) will dominate the error budget.  At higher accretion rates, the relationship may curve due to changes in the slope of the accretion continuum.

\begin{table}[]
    \centering
    \caption{Correlations between photometry and $\dot{M}_{acc}$}
    \label{table:fitting}
    \begin{tabular}{ccc}
    \hline
    \hline
    Photometry & Equation$^a$ & rms$^b$\\
    \hline
    TESS 2021 & $\dot{M} =1.45\times 10^{-8}\, T -1.16\times 10^{-8}$ & $7\%$ \\
    TESS 2023 & $\dot{M}=2.01\times 10^{-8}\, T-1.64\times 10^{-8}$ & $15\%$ \\
    ASAS-SN $g$ & $\dot{M}=4.90\times 10^{-9}\, g - 1.96\times 10^{-9}$ & $11\%$ \\
    \hline
    $g$ and TESS 2019 &  $g=3.12 T -2.10$ & $2.7\%$ \\
    $g$ and TESS 2021 &  $g=2.99 T -1.95$ & $2.8\%$ \\
    $g$ and TESS 2023 &  $g=3.57 T -2.40$ & $4.5\%$ \\
    $g$ and TESS 2025 &  $g=2.87 T -1.92$ & $4.0\%$ \\    
    \hline
    \multicolumn{3}{l}{$^a$Mass accretion rate  $M_{\odot}/\rm{yr}$, }\\
    \multicolumn{3}{l}{~~photometry T and $g$ unitless normalized flux }\\
    \multicolumn{3}{l}{$^b$Mean absolute percentage error}\\
    \end{tabular}
\end{table}

\subsection{Photospheric contribution to light curves }\label{section:photosphere}

 The photosphere contributes most of the flux in the TESS band and some of the flux in the $g$-band.  While TW Hya is expected to be spotted \citep[e.g.][]{debes13}, the pole-on orientation and stable magnetic orientation \citep{donati11,donati24,sicilia23} means that the fractional spot coverage and therefore visible photosphere is less variable than for many other young stars.  In this section, we use the correlations from Section \ref{section:correlation_tess}--\ref{section:correlation_gband} to infer the change in photospheric emission in the TESS bandpass and subsequently compare this value to measurements from spectroscopy.

The linear correlations with accretion rate exhibit a non-zero intercept (see Table \ref{table:fitting}), consistent with the presence of photospheric emission.  
From to the fitting equations presented in Table \ref{table:fitting}, a mass accretion rate (\(\dot{M}\)) of zero corresponds to a TESS normalized flux ($T$) of 0.80 in 2021 and 0.82 in 2023, or $\sim 9.65$ mag.  From the lowest veiling measurements from CHIRON, we estimate that $19\pm4\%$ (for 2021) and $13\pm3\%$ (for 2023) of flux in the TESS bandpass is produced in the accretion flow, consistent with the higher veiling in 2021 than 2023. The remaining flux ($\sim 81-87\%$) produced in the photosphere, which corresponds to $\sim 9.71$ mag and is relatively consistent with the measurement from linear fitting.

The $g$-band shows only a small offset between the photospheric flux estimated from spectra and that measured from correlations.  
From the right panel of Figure \ref{fig:linear_combined}, the photospheric normalized flux, at 0 accretion, is 0.40, corresponding to $\sim 12.5$ mag. However, the photospheric contribution in $g$-band is calculated empirically to be $g=12.35$ mag from low-resolution HST spectra.

\begin{figure*}[!t]
\centering
\includegraphics[scale=0.2]
{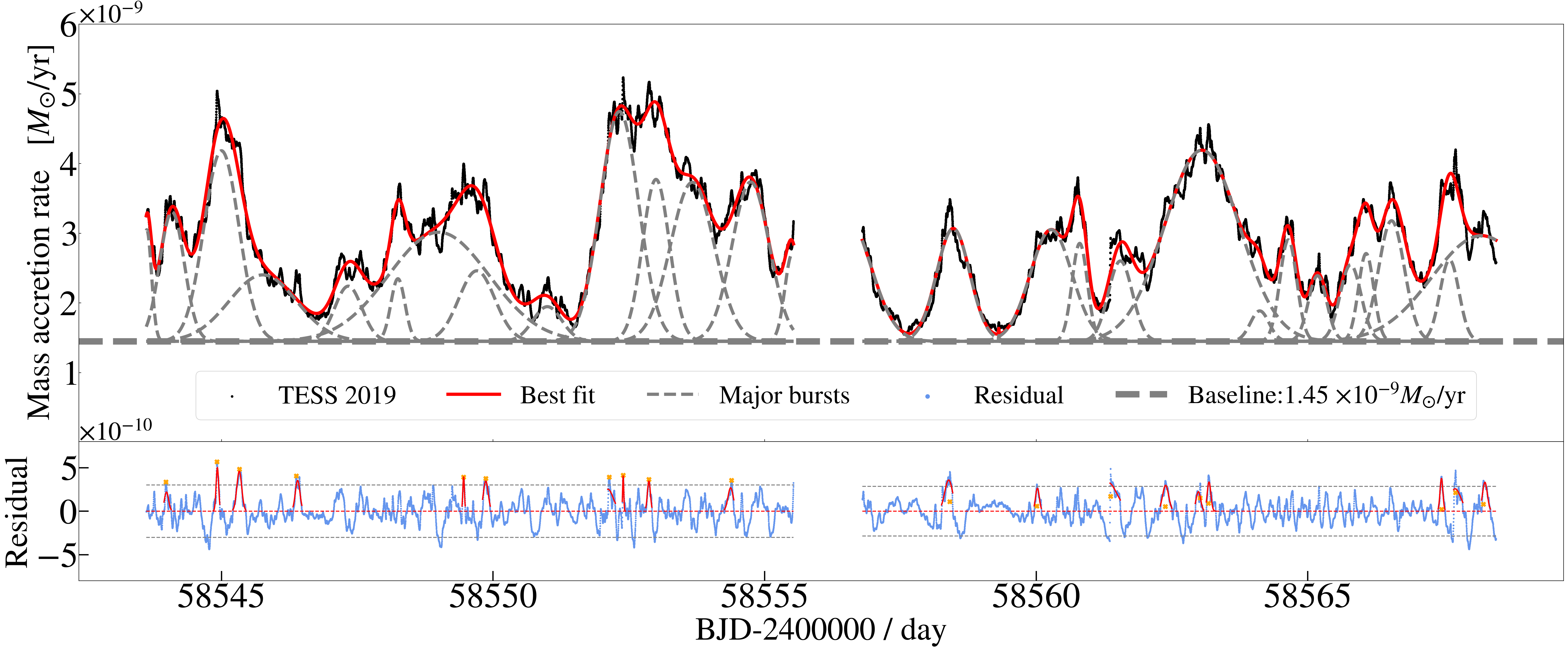}
\caption{Fits of Gaussian profiles (dashed gray, with a combined red curve) to bursts in the TESS 2019 accretion time series (black points).   The top panel shows major bursts, identified visually and measured as excesses of the baseline (dashed gray line) at the flux minimum during the epoch.
The bottom panel shows the residual from the fits to the major bursts (blue points).   The locations above $2\times$ RMS of the residual, identified by the orange crosses above the gray dashed line), termed minor bursts, are fit with Gaussian profiles to assess our sensitivity limits.}
\label{fig:tess2019_fitting}
\end{figure*}

The comparison of TESS and $g$-band flux in Figure \ref{fig:linear_combined} (middle panel) demonstrates that the assumption of a constant photospheric flux in the TESS band is reasonable, especially on short timescales.  The $g$-band normalized flux of 0.4, with no accretion, corresponds to TESS normalized fluxes of 0.78--0.81 in each year.  At a normalized TESS flux of 1, the four years have a standard deviation of 8\% from the best linear fits.  This value is higher than the standard deviations of 2.7-4.5\% differences within each year (Table~1).  These differences may be caused by long-term but small differences in visible spot coverage or by changes in the spectral slope of the accretion.

The spectroscopic estimates are calculated by comparing the equivalent width of isolated absorption lines from 6000--7000 \AA\ to high-resolution spectra of TWA 25, a weak-lined T Tauri star with an M0-M0.5 spectral type, similar to TW Hya \citep[e.g.][]{stelzer13,herczeg14}.  We then calculate the flux in the accretion continuum, assumed to be flat to long wavelength (see \citealt{herczeg23} for the spectral shape and discussion of templates), by convolving the template spectra with the TESS filter transmission curve.

\section{Measuring the time dependence in accretion rates}\label{section:results}

The TESS light curves provide us with changes on minutes-to-days timescales, while ASAS-SN light curves provide us with changes on month-to-year timescales.  In this section, we evaluate the full set of changes and search for variability timescales.  We begin by measuring the accretion bursts from TESS, then assess reset timescales, and finally assess long-term changes in accretion.

\subsection{Accretion bursts in TESS photometry}\label{section:gaussian fitting}
\subsubsection{Measuring bursts in the TESS lightcurves}

The TESS light curves can be modeled as a series of overlapping bursts, with a total range from $\sim1-9~\times~10^{-9}$\,M$_\odot$~yr$^{-1}$, consistent with the range of rates from \citet{herczeg23}. 
The correlations established in Section \ref{section:correlation} allow us to convert the TESS photometric light curves to a time series of mass accretion rates.  Each burst is approximated here as a Gaussian profile, characterized by burst duration and total accreted mass. While the bursts are not perfectly Gaussian, including several with rise times faster than decay times, Gaussian distributions provide a simple, uniform, and easily applied approximation to the observed profiles.

The measurements of bursts depend on initially estimating a quiescent level of quasi-constant accretion.  The four TESS light curves each exhibit a different baseline accretion rate. 
We roughly estimate the baseline accretion rate by taking the minimum value in each TESS observation. 
The baseline accretion rate for TESS 2025 of $1.83\times 10^{-9}$ M$_\odot$ yr$^{-1}$ is significantly higher than the other three baselines ($1.45\times 10^{-9}$ M$_\odot$ yr$^{-1}$ in 2019, $4.69\times 10^{-10}$ M$_\odot$ yr$^{-1}$ in 2021, and $6.58\times 10^{-10}$ M$_\odot$ yr$^{-1}$ in 2023).

\begin{figure*}[!t]
\centering
\includegraphics[scale=0.6]
{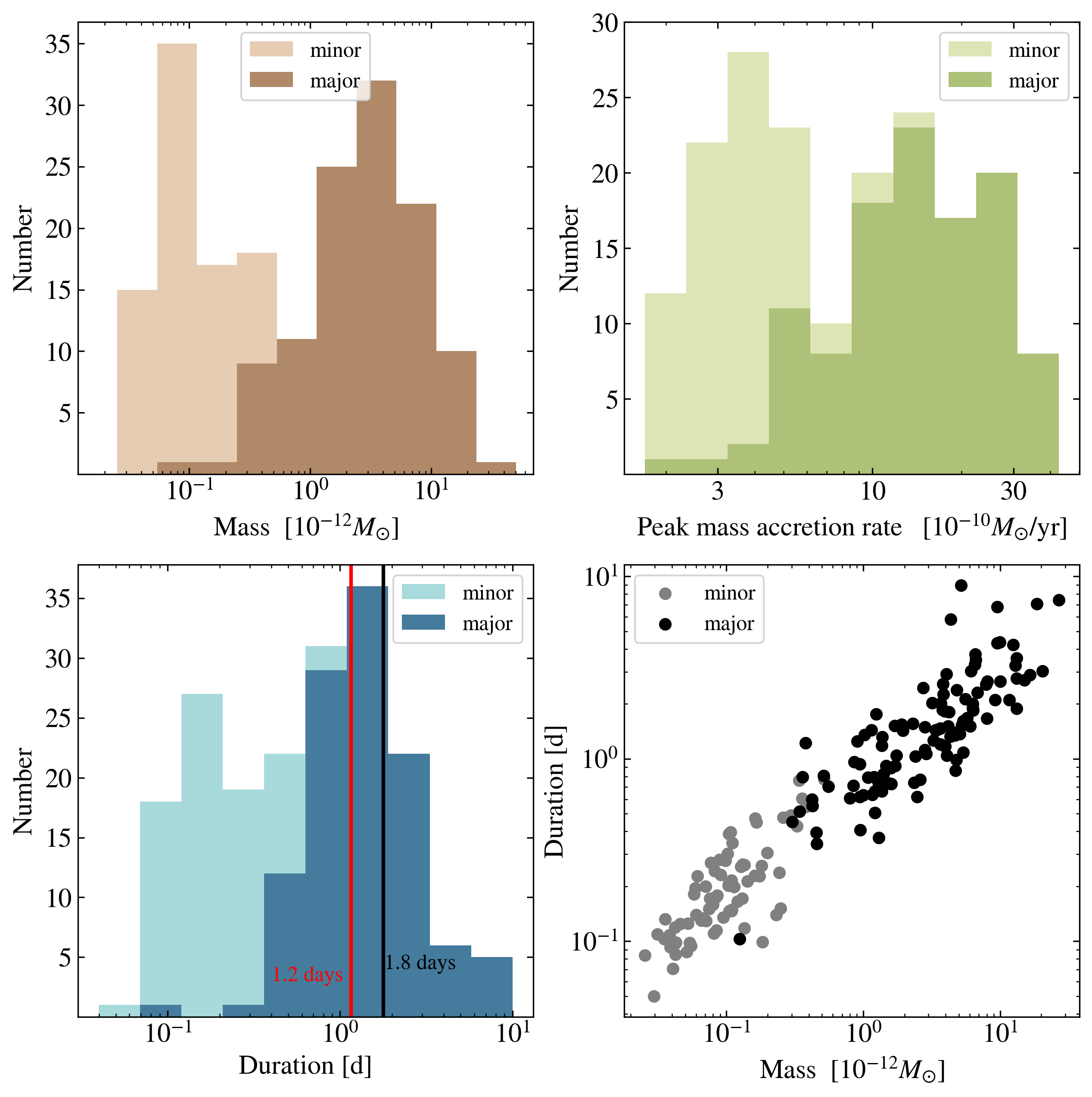}
\caption{Summary of the properties of major and minor bursts measured from model fits, including burst mass (upper left), peak mass accretion rate (upper right), duration (lower left), and mass versus duration (lower right),  In each plot, the darker regions or points show the bursts while the lighter shades show our sensitivity limits from fits to the residual time series, termed minor bursts.}
\label{fig:mass_duration_combined}
\end{figure*}

Bursts are measured from multi-Gaussian fits to the mass accretion time series, after subtracting off the estimated baseline. 
The time series are fit with multiple Gaussian profiles using the Python software lmfit \citep{matt24}. For each TESS light curve, we first use multiple Gaussian profiles to obtain the best fitting result of the largest bursts, called {\it `major bursts'}, as identified by eye. The total mass of each burst is calculated as the area $A$ of the Gaussian component, while the duration is defined as twice the full duration at half maximum (Duration = $2 \times$ FWHM$=4\sqrt{2\ln{2}}\sigma$) of the Gaussian component.  Though the identification of bursts is not unique, they provide a reasonable interpretation of the light curves. 

After subtracting off our best fits to the major bursts, the residual includes some significant fluctuations that are similarly fit with Gaussian profiles.  We term {\it `minor bursts'} as peaks above $2\times$ root mean square (RMS) of the residual.  These minor bursts are expected to be a combination of real bursts and false bursts that are introduced by imperfect fits of Gaussian profiles to the major bursts. 
The properties of these minor bursts provide an estimate of our sensitivity limits.

To illustrate our approach, Figure~\ref{fig:tess2019_fitting} shows the fit of the TESS 2019 mass accretion time series, including the major and minor bursts. The details of our fitting method are described in the Appendix \ref{appendix:gaussian}. The initial light curve has a scatter of $3.0\times~10^{-9}$ M$_\odot$~yr$^{-1}$.  After the major bursts are subtracted from the time series, the scatter is reduced to $1.5\times 10^{-10}$ M$_\odot$~yr$^{-1}$. Finally, after fitting to the minor bursts, the scatter in the light curve is $1.1\times~10^{-10}$~M$_\odot$~yr$^{-1}$. This approach is repeated for the TESS light curves obtained in 2021, 2023 and 2025 (see figures in the Appendix \ref{appendix:gaussian}). The four TESS light curves include a total of 112 major bursts: 28 in 2019, 25 in 2021, 26 in 2023, and 33 in 2025. 

\subsubsection{Properties of the bursts}

The burst peak, total accreted mass, and duration all depend on the baseline accretion rate.  We tested how burst properties would change with baseline accretion rates of  $20 \%$, $50 \%$ and $80 \%$ of the minimum accretion rate in each TESS observation. Lower baseline values lead to the identification of more and longer bursts with higher peak accretion rate in our fitting, which risks obscuring the true burst signals originating from the mass accretion process. For simplicity, we ultimately assume the minimum accretion rate within each TESS observation as the baseline accretion level.

The mass distribution of all bursts (upper left panel of Figure \ref{fig:mass_duration_combined}) ranges from $10^{-13}-10^{-11}$ M$_{\odot}$.   
The peak mass accretion rate of all the bursts (upper right panel of Figure \ref{fig:mass_duration_combined}) range from  $1-3 \times 10^{-9}$ M$_{\odot}$~yr$^{-1}$.  
The duration distribution of all bursts (lower left panel of Figure \ref{fig:mass_duration_combined}) show that bursts have an average duration of 1.8 days.  Most burst last for 1--2 days, with a maximum duration of $\sim 9$ days.  The long bursts may simply be multiple bursts that lead to broad profiles in the light curve.  The total mass and duration of bursts are strongly correlated (lower right panel of Figure~\ref{fig:mass_duration_combined}).

The minor bursts demonstrate our sensitivity limits.  These minor bursts are a mix of real accretion bursts, short flares, and residuals from fits to major bursts.
Most of the minor bursts last for 0.1--0.3 days, with peak fluxes of $3-5 \times 10^{-10}$ M$_{\odot}$~yr$^{-1}$ and a rough sensitivity limit on total mass of $10^{-13}$ M$_{\odot}$. 
This sensitivity limit of $\sim 10^{-13}$ M$_{\odot}$ corresponds to emitted energies of $\sim 10^{35}$ erg.  For accretion, this energy would equate to the potential energy of the infalling gas at the magnetospheric truncation radius.  This same energy can be produced by strong chromospheric flares produced by magnetic activity, so this empirical limit in our measurements also sets a sensitivity level where magnetic activity may be mistaken for accretion bursts (see for example the discussion of flares versus accretion bursts in \citealt{tofflemire17} and description of flares in \citealt{stelzer22}). \citet{siwak18} also found minor bursts, with durations of 11-30 min and amplitudes of 0.1-0.5, superimposed on MOST and ground-based lightcurves of TW Hya.
Our identification of bursts is not sensitive to long, low-amplitude changes because the light curve is characterized by constant bursting and the TESS time coverage in a single visit is insufficient to identify sustained changes in brightness.

\subsection{Timescales from structure functions}\label{section:structure function}
Structure functions provide timescales for changes by comparing the relative values of all data points versus the difference in time.  Structure functions were first applied in astronomy to AGN \citep[e.g.][]{devries05,schmidt10} and have since been applied to accretion onto young stars \citep[e.g.][]{rigon17,sergison20,herczeg23,zsidi25}.

We
apply the structure function to the TESS and ASAS-SN photometry to measure a relatively robust non-periodic variation timescale of TW Hya. For every two data points in our data set, the difference in target magnitude ($|\Delta m|$) is compared with the time difference ($|\Delta t|$). A steady increase in $|\Delta m|$ with respect to $|\Delta t|$ indicates that the system moves further away from the initial point as time passes, which implies a system in a dynamic state. In contrast, a constant trend of $|\Delta m|$-$|\Delta t|$ implies a stable system with shorter-term variations.  For both the TESS and ASAS-SN structure functions, we adopt a bin size of $\Delta(\log t) = 0.033$, following the choice of \citet{herczeg23}.  This bin size corresponds to ∼30 bins per dex and is simply a practical choice for smooth sampling in $\Delta \log t$.

\begin{figure*}
\centering
\includegraphics[scale=0.25]
{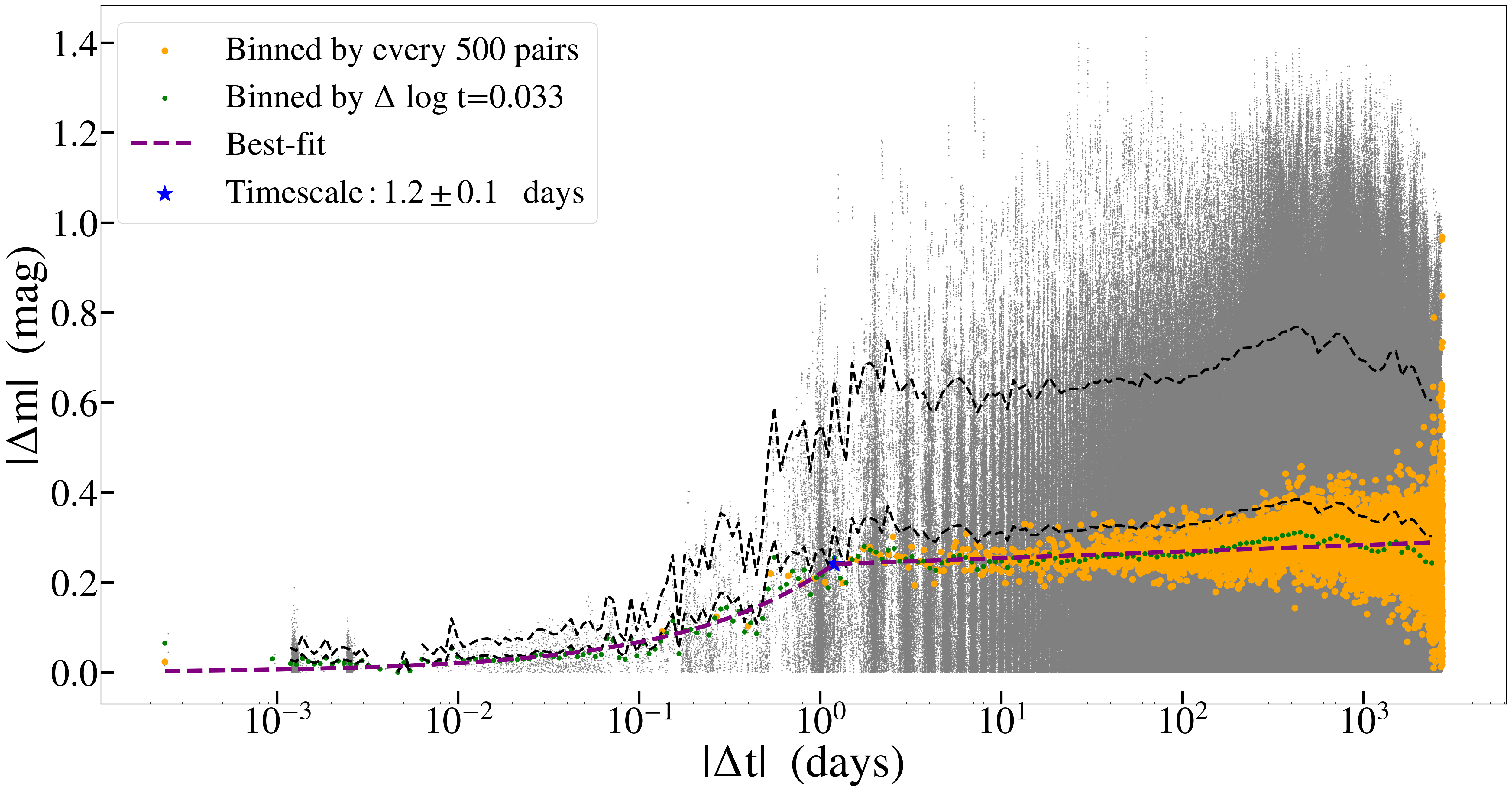}
\caption{The structure function ($|\Delta m|$-$|\Delta t|$ distribution) for 
ASAS-SN $g$-band photometry for 10240075 pairs from the 4526 $g$-band measurements (gray points). The median $|\Delta m|$-$|\Delta t|$ in bins of every 500 pairs of $|\Delta t|$ are presented in orange points and binned by $\Delta \log t = 0.033$ shown in green points. 
A two-part function (Table \ref{tab:fitting_paras}) is fit to the time-binned data (purpled dashed line), with reset timescale marked by the blue star.  
The 1-sigma and 2-sigma distributions (black dashed lines) indicate the regions that include approximately $68 \%$ and $95 \%$ of the data points.}
\label{fig:structure_function_asassn}
\end{figure*}

\subsubsection{Reset timescale for ASAS-SN $g$-band photometry}\label{section:structure_function_asassn}

\begin{table*}[!t]
    \centering
    \caption{Two-part fits to Structure Functions}
    \label{tab:fitting_paras}
    \begin{tabular}{lccccc}
    \hline
    \hline
    Parameters & ASAS-SN $g$-band & TESS 2019 & TESS 2021 & TESS 2023 & TESS 2025 \\
    \hline
    $A_0$ & $0.220 \pm 0.007$ & $0.157 \pm 0.003$  & $0.185 \pm 0.004$ & $0.156 \pm 0.002$ & $0.115 \pm 0.002$ \\
    $\beta $ & $0.51 \pm 0.02$ & $0.55 \pm 0.02$  & $0.63 \pm 0.03$ & $0.56 \pm 0.01$ & $0.58 \pm 0.02$\\
    $A_1$ & $0.014 \pm 0.002$ & $-0.025 \pm 0.004$  & $-0.09 \pm 0.01$ & $-0.012 \pm 0.003$ & $-0.003 \pm 0.002$\\
    $t_0$ & $1.2 \pm 0.1$ & $1.2 \pm 0.1$  & $2.0 \pm 0.1$ & $1.36 \pm 0.05$ & $1.9 \pm 0.1$ \\
    \hline
    \end{tabular}
\end{table*}

Figure \ref{fig:structure_function_asassn} shows 
the structure function of ASAS-SN $g$-band measurements.
The median $|\Delta m|$ increases consistently with $|\Delta t|$ on a relatively short time scale (< 2 days). However, on longer timescales (> 2 days), the binned value becomes much flatter with respect to $|\Delta t|$ and is stable ($|\Delta m|$ for the g-band stays at about $0.3$ mag).

To measure the precise reset timescale (listed in Table \ref{tab:fitting_paras}), we fit the median binned $|\Delta m|$-$|\Delta t|$ relations (in green points) with a two-part function in the following, 
\begin{equation}
\label{eq:two-part}
|\Delta m| = 
\begin{cases}
  A_0 |\Delta t|^\beta,  \& |\Delta t| < t_0 ;\\    
  A_1 \log \left( \frac{|\Delta t|}{t_0} \right) + A_0 t_0^\beta, \& |\Delta t| \ge t_0, 
\end{cases}
\end{equation}
in which $A_0$, $\beta$, $A_1$, and $t_\mathrm{0}$ are set as free parameters. The unit of $|\Delta m|$ is magnitude, and the units for $|\Delta t|$ and $t_0$ are days in the equation.

The $g$-band structure function has a reset timescale of $t_0 = 1.2 \pm 0.1$ days, similar to the average duration~$1.8$ days measured in the lower right panel of Figure \ref{fig:mass_duration_combined} and to the results from \citet{herczeg23} for spectroscopy.  The sampling of the ASAS-SN data has a minimum interval of 0.0002 day for the $g$-band and a time baseline of about 2700 days.  These intervals meet the requirements from \cite{devries05} that the timescale in the structure function is reliable if the time value is $\sim 30$ times greater than the sampling interval and $\sim 1/15$ of the total time baseline.

\subsubsection{Reset timescale for TESS photometry}
We apply the same $|\Delta m|$-$|\Delta t|$ analysis separately to the four TESS light curves, with fitting results of the two-part function in the Appendix \ref{appendix:structure_function}. Given the high-precision TESS data and the large number of paired points ($\sim 1.5\times10^{8}$ pairs for 2019, 2023, and 2025 and $\sim 6\times10^{6}$ pairs for 2021),
we provide a 2D hexagonal binning plot (a type of two-dimensional histogram that uses hexagonal cells to visualize the density of data points), time binning plot, and the fitting result. 

The reset timescales range from 1.2--2 days, consistent with the previous measurements of 1.2 day timescales from sparser photometric and spectroscopic observations (see Section \ref{section:structure_function_asassn}).  The negative slope in the structure function for TESS 2019 and 2021 is a consequence of the short TESS observing windows.  The TESS datasets have time coverage of 25 days, at the limit of the duration ($1.2-2.0$ days of reset timescale $\times 15 = 18-30$ day observing period) that is required for reliability \citep[see the criteria developed by][]{devries05}.

\subsection{Searching for extended periods with weak or strong accretion}
\label{section:long_variability}

The structure function from the ASAS-SN $g$-band data flattens at $\sim 1.2$ days at a level of $\sim 0.3$ mag.  This flattening suggests that variability on longer timescales is insignificant.  However, we visually identify some long-term differences in the $g$-band photometry in Figure~\ref{fig:asassn}, and in this section we quantify the significance of these changes.

To assess the level and significance of long-term variability, we measure the average magnitude in each visibility window and in 30-day segments.  We then compare those distributions to bootstrapped distributions, where each photometric data point is randomly reassigned to a different observation time.  Bootstrapped distributions are calculated 1000 times, and only time windows with at least 25 data points are included in this analysis.

We find significant but modest long-term variability.  In 30-day intervals, the brightness ranges from $g=11.20-11.76$ with a standard deviation of 0.13 mag, larger than the randomized standard deviation of 0.03 mag. 
The month-long fluctuations are listed as $g$-band brightness and accretion rates in Table \ref{tab:averages} in Appendix \ref{appendix:long-term}.

When averaged over the visible annual windows, the the brightness ranges from $g=11.33-11.65$ mag with a standard deviation of 0.10, far above the standard deviation of 0.01 mag expected from bootstrapping. 
From the start of ASAS-SN monitoring in Dec.~2017 until June~2020, TW Hya had an average accretion rate of $\sim 3\times10^{-9}$ M$_\odot$ yr$^{-1}$.  The accretion rate then dipped to $\sim 2\times10^{-9}$ M$_\odot$ yr$^{-1}$ in early 2021, before rising up to $\sim 4\times10^{-9}$ M$_\odot$ yr$^{-1}$ in late 2021 and then returning to a level slightly below average.

Spectroscopic veiling measurements from \citet{herczeg23} demonstrate that these long-term trends in $g$-band brightness are dominated by accretion variability.  In the bright g-band epoch, from HJD 59500--59750 with $g=11.3$, the average veiling (at 5000-5100 \AA) is 0.86 (median 0.90) from 42 CHIRON spectra.  During the faint g-band epoch (HJD 59200-59400, with $g=11.65$), the average veiling is 0.46 (median of 0.34) from 26 CHIRON spectra. This change is consistent with expectations for enhanced accretion. 
If the brightness change in $g$-band was caused by spots, then the photosphere would be roughly 1.4 times brighter and the veiling would have been lower during the brighter epochs.  \citet{siwak18} and \citet{wendeborn24} reach similar conclusions based on the variability being nearly achromatic in multi-band monitoring.
However, we expect some minor contributions of changing spot coverage to the variable optical brightness of TW Hya \citep[see, e.g.][]{espaillat19}, as well as short chromospheric flares \citep[e.g.][]{siwak18}.
Some of the $g$-band fluctuations could also be generated by a change in the accretion spectrum, if more emission escapes shortward of the Balmer jump during periods when the $g$-band is fainter.

\section{Discussion} \label{section:discussion}

TW Hya is an old (10-20 Myr, see discussions in \citealt{luhman23} and \citealt{herczeg23}) Classical T Tauri star with a gas-rich disk and ongoing accretion.  The optical spectrum is a combination of emission from the photosphere and the accretion flow and shock \citep[e.g.][]{calvet98}.    
Our study provides the first systematic measurement of the mass distribution and duration of short accretion bursts by calibrating photometry to simultaneous accretion rate measurements.   We also establish that seasonal variations are modest but significant.

We discuss separately the implications for short- and long-term variability, including implications for the cooling timescale, and then discuss how the framework for TW Hya may be applied as a probe of accretion onto other young stars.

\subsection{The interpretation of optical variability}

In the standard accretion shock paradigm of \citet{calvet98}, the pre-shock gas is optically thin and dominates the Balmer continuum emission.  At longer wavelengths, including the ASAS-SN $g$-band and TESS, the emission is dominated by the Paschen continuum and H$^-$ continuum, produced mostly in the post-shock gas and heated photosphere.  The accretion flow may have multiple components, in which case the lower density flows would contribute more to the red emission \citep[e.g.][]{espaillat19,pittman25}.

The bursts seen in TESS  must be changes in the measured amount of the Paschen and H$^-$ continuum emission.  The visible emission from the hot spot changes quickly, with rise and decay times typically a few hours.

Our interpretation is that these bursts trace changes in the actual accretion rate, even though we are only measuring changes in the hot spot.  In Section \ref{section:thermal}, we explore further how quickly the accretion flow may heat and the hot spot may cool.  This interpretation is supported by the strong correlation of the TESS brightness with simultaneous veiling measurements.  The veiling measurements themselves are also a measure of the Paschen and H$^-$ continuum, but they are correlated with many lines, including  \ion{He}{1} lines and \ion{He}{2} $\lambda4686$, that should directly trace the accretion shock \citep[e.g.][]{beristain01,yang07}.

There may be alternative explanations to apparent bursts in light curves, including hot spots that rotate across the star, leading to variations in their 
visible projected surface area \citep[see, e.g., the case of RU Lup,][]{armeni24}.  While wavelet and periodogram timescales roughly correspond to the stellar rotation period, TW Hya is viewed nearly pole-on, so only equatorial accretion hot spots could rotate behind the star.  In the simulations of \citet{romanova25}, the unstable accretion regime leads to sporadic accretion in tongues that start as equatorial but are then channeled by higher order magnetic fields to latitudes of 35-45$^\circ$.  Although the phase-folded light curves are not sinusoidal, phase-folded light curves of unstable spots at different latitudes could reproduce at least some of the quasi-periodicity and apparent bursts \citep{armeni24}.

 If the bursts are interpreted as a change in hot spot emission and not simply a geometrical effect, then the measurements of bursts are robust  whether the accretion primarily flows to the poles or whether the burst may include mid-latitude accretion.  In the Zeeman Doppler Imaging maps, the accretion flow strikes near the pole with a relatively stable morphology \citep{donati11,donati24,sicilia23}, consistent with speculation from increased line emission at high velocities that mass loading during bursts occurs along those dipole-like lines \citep{herczeg23}.
In the \citet{romanova25} simulations the accretion tongues strike at mid-latitudes.  Either way, the detected emission is produced in accretion hot spots.  The accretion tongues may have smaller truncation radii, which would change the initial potential energy and therefore the final luminosity by 10-20\%.

If the heating and cooling timescales of the hot spot were longer than burst timescales, then the emission from the heated photosphere would be expected to be stable.  The variability would then be caused by changes in only the pre-shock emission.  In this case, we would be underestimating the mass in bursts while overestimating the quiescent accretion rate. We also assume that the baseline accretion rate is roughly constant over short timescales, although the quiescent rate may itself be the superposition of frequent smaller bursts.

\subsection{Timescales and rotation of the star and inner disk}

Extensive analyses of light curves of TW Hya, initially with MOST \citep{rucinski08,siwak14,siwak18} and most recently with ASAS-SN and TESS \citep{sicilia23,wendeborn24,koen25}, has established that its accretion flow is bursty, consistent with light curves of most other accreting young stars \citep[e.g.][]{stauffer14,cody18}.  We find that the base accretion rate is typically $\sim 1 \times 10^{-9}$ M$_\odot$ yr$^{-1}$, but the frequency of bursts means that usually TW Hya is accreting at a higher rate.  The peak rate found during the four TESS integrations is 
$\sim 9\times 10^{-9}$ M$_\odot$ yr$^{-1}$.  The average accretion rate from the ASAS-SN is usually a factor of a few above the baseline (see Table \ref{tab:averages}).  More mass arrives during these short bursts than during the baseline quiescent periods.

The timescale for changes can be described in several different ways.  For clarity, we step through these different timescales:

\begin{description}

\item[Stellar rotation period]  The rotation period of 3.6 days is measured from sinusoidal variations of the radial velocity \citep[e.g.][]{huelamo08}.

\item[Quasi-periodic timescales]  Periodogram analyses of light curves lead to timescales (or quasi-periods) of 3.5-4 days \citep[e.g.][]{wendeborn24,romanova25}.  These quasi-periods are measurements of when bursts occur.  While these bursts may be related to rotation, the rotation period should not be measured from non-sinusoidal photometry.

\item[Burst durations and frequencies]  Accretion bursts are characterized by their duration, frequency, and amplitude.

\item[Reset timescale]  The reset timescale is the timescale over which accretion rates are independent and unrelated.  Structure functions lead to reset timescales of 
1.2-2 days (see Figure \ref{fig:structure_function_asassn} and Table \ref{tab:fitting_paras}), consistent with the reset timescale of 1.6 days found with spectroscopy \citep{herczeg23}.  A pure sinusoid would have a reset timescale of half the period\footnote{Technically the calculation of a reset timescale for a sinusoid breaks down, but the initial peak of the structure function occurs at half of the period.}, so this timescale is also consistent with the rotation period.
\end{description}

The burst properties and associated timescales all paint a similar picture.  Bursts occur quasi-periodically and typically last $\sim 1$ day.  The magnetospheric radius of 3.5 R$_*$ \citep{gravity20} to 4.5 R$_*$ \citep{donati24} leads to orbital periods of 1.3--2 days.  Since the magnetic field rotates with the star on a period of $\sim 3.6$ days \citep{donati24}, the disk truncation radius is interior to co-rotation radius.  The burst frequency of $\sim 1$ per day may suggest that bursts occur more often for TW Hya than for the broader sample studied by \citet{stauffer14}, who found 0.2 bursts per day. However, this apparent discrepancy may in part be attributed to differences in methodologies, including in peak finding algorithms and duration requirements.  Geometric differences are also expected because we never see the back side of TW Hya.

These different timescales described above are consistent with regular, but not strictly periodic, bursts \citep[e.g.][]{armeni24}.  In this bursty regime, unstable magnetic modes may load mass onto different accretion
tongues, as simulated by \citet{romanova25}. 
Considering the similarity between the reset timescale and the orbital timescale of $\sim 1.3$ days at the inferred magnetospheric radius from Br$\gamma$, the disk structure seems to be smeared out by the orbital motion. In other words, substructures at the truncation radius from unstable accretion cannot survive for more than one rotation period.

The stochastic bursts may also be associated with the accretion of individual turbulent cells that form in an accretion disk due to magnetorotational instability \citep[MRI;][]{balbus98}. The MRI is likely not active across the full disk, since most of the gas in a disk is cold (typically tens of K), weakly ionized, and only weakly coupled with the magnetic field \citep[see, e.g.,][]{bai13}. 
However, the innermost disk is likely warm enough to support the MRI \citep{ueda22}, which would drive turbulence \citep{romanova12,cecil24}.  In this case, matter may reach the inner disk at a variable rate, leading to variations in the disk-to-star accretion rate.

\subsection{Thermal response time of the hotspot versus the burst duration}\label{section:thermal}
We interpret the variability observed in TW Hya as caused by accreting clumps of disk material, which are loaded onto stellar magnetic field lines and funneled to the surface. In this scenario, gravitational potential energy is converted to kinetic energy during infall and is then converted to thermal energy upon impact near the stellar photosphere. The thermal energy is deposited in a shocked region and then dissipated and reprocessed, resulting in excess emission from the shocked region, the pre-shocked in-falling gas, and a localized hotspot on the stellar photosphere \citep[e.g.][]{calvet98, Muzerolle01}.

A burst can therefore be thought of in phases. First excess gas enters the magnetospheric funnel flow.  The increased flow then shocks at the stellar surface and heats the layers beneath the shock. Those layers adjust their temperature and spread the excess energy, and finally the system radiates the energy away.

In a quasi-steady state, the outgoing radiative flux from the hotspot balances the inflowing kinetic energy flux from accretion. When the accretion rate changes, the hotspot requires some timescale to reach the new equilibrium, here termed the hotspot thermal response time, as follows:
\begin{equation}
    \tau_{\rm th}=\frac{\frac{3}{2}NK_BT_0}{\frac{1}{2}\dot{M_0}v^2_{\rm ff}},
\end{equation}
where $\dot{M_0}$ is the baseline accretion rate, $v_{\rm ff}$ is the free fall velocity, $T_0$ is the temperature of the hotspot at the equilibrium.
We explored the thermal response of the hotspot using a reservoir model, in which the heated photosphere is characterized by the number of particles that are participating over the photosphere heating process. This number is set by the effective density, depth and area of the hotspot under simple assumptions about photospheric density and geometry of the hotspot. For TW Hya, we adopt photospheric densities from PHOENIX atmosphere models \citep{Husser13}, 
appropriate for its stellar parameters 
($T_{\rm eff}=3800$\,K and $\log(g)=4.2$, \citealt{sokal18}),
which yield values of $\rho \sim 5\times10^{-7} - 10^{-6}$ \,\, [$g$\,cm$^{-3}$] near optical depth $\tau\approx 1$.

We obtain thermal response times ranging from 5 minutes to 2.6 hours for reservoir depths from one photospheric scale height ($\sim 500$ km) to 1$\%$ stellar radius ($\sim 6470$ km) and a typical filling factor of 1$\%$ for TW Hya. In order for the response time to approach $\sim\,0.5-1$ day, the reservoir would need to extend to a depth of $h \sim 0.1\,R_\ast \approx 65000$ km, which appears implausible because the stellar-interior temperature increases rapidly with depth, inhibiting the downward flow of heat.

If the hotspot thermal response time is shorter than the timescale of accretion variability, then the system stays close to equilibrium and the emergent luminosity follows the instantaneous accretion rate. Conversely, if the response time were longer, a measurable lag would appear between accretion changes and observed brightness. Our estimates suggest response times of minutes to a few hours.  
Since this response time is much shorter than the day-long burst durations, the photometric variability timescales correspond to the actual timescales of mass impacting the star in variable clumps.
Nevertheless, detecting or constraining such short lags remains an observational challenge. Simultaneous high-cadence photometry and spectroscopic accretion measurements could directly probe the thermal inertia of the hotspot through the accretion shock \citep[e.g.][]{dupree12}, offering a novel diagnostic of the conditions in the heated photosphere.

Previous numerical models of accretion shocks have mainly focused on the structure and emission from the post-shock region at X-ray wavelengths \citep{Sacco08,Drake09,Orlando10,Matsakos13}. These studies demonstrate that the shocked gas undergoes strong radiative cooling and may experience thermal instabilities, but they generally do not address the subsequent heating or adjustment of the stellar photosphere itself.  In contrast, \citet{calvet98} linked the post-shock emission to the radiative equilibrium of the underlying photosphere, providing the basis for interpreting the optical and UV excess as emission from a heated stellar surface. Our simplified treatment builds on that framework to estimate the characteristic thermal response of these deeper layers and explore their possible time-dependent behavior.

\subsection{Variability on longer timescales} \label{sec:longer}

We find long-term variations in month-long averages of $g$-band brightness ranging from 11.20--11.80, with an average of $11.51$ and a standard deviation of $0.13$\,mag. These changes correspond with accretion rate changes from $1.7 \times 10^{-9}$ to $4.3 \times 10^{-9}$\,M$_\odot$~yr$^{-1}$.  Annual averages, calculated for each visibility window, still have a standard deviation of $0.10$ mag.  
Persistent low and high periods were previously noted from spectroscopic measurements but were not analyzed for significance by \citet{herczeg23}. 

The periods of apparently stronger and weaker accretion may occur because of a change in the base accretion rate or by increasing or decreasing the strength and frequency of bursts.    The 2025 TESS light curve has a higher baseline than other years, with bursts that have lower peak/quiescent ratios than other years.  Based on this comparison and the measured change in veiling, the long-term fluctuations in accretion rate are likely caused by a change in the baseline rate rather than the bursts. 

Long-term variations on timescales of 100 days, as likely seen here, could relate to changes in stellar magnetic activity modulating accretion rate. For solar-type main-sequence stars, starspot lifetimes are $\sim 100$ days \citep{namekata19}.  Young stars are more heavily spotted than older stars \citep[e.g.][]{cao22}, with spot structures that evolve on year-long timescales \citep[e.g.][]{grankin08,gully17}.  Any reorganization of the stellar magnetic geometry, as manifested in starspots, may occur on $\sim 100$ day timescales.  These changes could then affect the accretion rate in some way, for instance in the location where the dipole-like field intercepts the disk.  These variations may also be a change in visible spot coverage instead of accretion.  Spot coverage is likely to fluctuate over months and years.  A dedicated spectroscopic analysis would be required to assess whether these changes are associated with spots.

There are no periods of severely depressed accretion.  Simulations by \citet{zhu24} formed long-lasting magnetic voids, but such voids have not been detected in the innermost disk of TW Hya in the past decade.

\subsection{A framework for interpreting accretion bursts onto other stars} \label{sec:framework}

The specific properties of TW Hya allow us to reliably convert TESS photometry to accretion rates.  Because the accretion flow onto TW Hya is viewed nearly pole-on, excess dust does not pass in our line-of-sight, which in other sources would cause extinction changes.
The four TESS light curves do not include any measurable spot signature, consistent with the expectation that the visible spot fraction is stable over short timescales.  

This framework may be applied to other accreting YSOs with similar characteristics.  
 The large set of TESS light curves of classical T Tauri stars \citep[e.g.][]{robinson22,serna24} include many other sources that do not suffer from short-term spot variability or extinction events.  The set of accretion burst masses, durations, and frequencies could then be used to test simulations of magnetospheric accretion in different magnetic regimes.  Similar approaches have been implemented to demonstrate enhancements of accretion rate in close binaries at periastron \citep{tofflemire17b,tofflemire25}.

 Conversions from photometry to accretion luminosity and accretion rate are necessarily unique to each star.  For the case of TW Hya, the conversion was developed with extensive measurements of accretion rate that were obtained simultaneously to some of the TESS monitoring. 
  Multi-band photometric datasets, especially including accretion sensitivity with a $u$-band filter \citep[e.g.][]{venuti21,wendeborn24}, must be planned in advance and have the multiple bands obtained nearly simultaneously to benchmark larger, single-band datasets.
The ASAS-SN monitoring also provides a massive dataset to understand variability over years and can be leveraged alongside TESS.

We interpret the changes in $g$-band brightness as related to accretion, although there may be contributions from variations in the spectral shape of the accretion continuum and in the spot coverage.
Ultimately, multi-band photometry or spectroscopy is required to have complete confidence that seasonal changes are the consequence of changes in accretion rate rather than accretion or spot structure.  The structure functions from TESS and ASAS-SN both show that simultaneity within $\sim 2$ hrs is usually sufficient for most applications.
Moreover, short brightening events, especially including some of the minor bursts, may be chromospheric flares rather than accretion bursts \citep[e.g.][]{tofflemire17,stelzer22}.

\section{Summary}
\label{sec:summary}

We reassess the photometric variability of TW Hya with long-term ASAS-SN $g$-band and high-time-resolution TESS photometry. We convert the photometric variability to mass accretion time series to allow the first time systematic measurements of the mass and duration for the short bursts that occurred in the accretion onto TW Hya. We also measure reset timescales with structure functions. From analyzing the TESS and ASAS-SN data, we find:

\begin{enumerate}
    \item The photometric variability is strongly correlated to the mass accretion rate, with tight linear relationships. Much of the short-time photometry variability comes from the accretion process. The slopes of fitting from different datasets differ, which could be explained by changes in visible spot coverage, temperature and density distribution of accretion flow.  The flux values where there is no accretion is approximately consistent with measurements of photospheric emission from spectra.

    \item We fit each burst in TESS accretion time series with Gaussian profile and provide the first systematic measurement of the mass and duration for 112 bursts. The major bursts have a total mass ranging from a sensitivity limit of $10^{-13}$\,M$_\odot$ to $3\times 10^{-11}$\,M$_{\odot}$.  For physical intuition, TW Hya accretes the mass of one large comet, though this mass should be interpreted as the size of the instability rather than a physical object.

    \item From the burst duration distribution, we measure an average duration of $1.8$ days for the major bursts. These timescales exceed the reasonable thermal response timescales of hotspots, which are on the order of minutes to hours, indicating that the bursts are likely driven by instabilities at the magnetosphere-disk interface. We also measure reset timescales ranging 1.2--2 days from structure functions of both ASAS-SN $g$-band and TESS photometry, roughly consistent with  the 1.6-day reset timescale in \cite{herczeg23}. The accretion process is turbulent  and fluctuates a lot on day-timescale.
    \item In the long term, TW Hya also shows significant variability. In 30-day intervals, the brightness ranges from $g=11.20-11.76$ with a standard deviation of 0.13 mag, larger than the randomized standard deviation of 0.03 mag.  The long-term fluctuations are caused by accretion rate change, as seen from spectroscopic veiling measurements from \citet{herczeg23}.
    \item Our method of converting TESS photometry to mass accretion rate could also be applied to other accreting YSOs with similar characteristics as TW Hya. A complete set of accretion burst masses and durations could potentially be used in future simulations, helping us understand the variable accretion process better.
\end{enumerate}

\section{Acknowledgements}

We thank the anonymous referee for a careful and rigorous read, which led us to re-evaluate some assumptions and resulted in a strengthened paper.
We thank the ODYSSEUS team \citep{espaillat22} for providing the framework that helped to motivate this paper.
  We also thank Beate Stelzer, Zhen Guo, Joel Kastner, for helpful comments on the manuscript.  JS thanks Nuria Calvet and Lee Hartmann for valuable discussions on the heated photosphere.

G.J.H. is supported by the general
grant 12173003 from the National Natural Science Foundation
of China and by National Key R\&D
program 2022YFA1603102 from the Ministry of
Science and Technology (MOST) of China.
D.J.\ is supported by NRC Canada and by an NSERC Discovery Grant.

This paper includes data collected with the TESS mission, obtained from the MAST data archive at the Space Telescope Science Institute (STScI). Funding for the TESS mission is provided by the NASA Explorer Program. STScI is operated by the Association of Universities for Research in Astronomy, Inc., under NASA contract NAS 5–26555.
This research has made use of the Spanish Virtual Observatory (https://svo.cab.inta-csic.es) project funded by MCIN/AEI/10.13039/501100011033/ through grant PID2020-112949GB-I00.  The TESS data presented in this article were obtained from the Mikulski Archive for Space Telescopes (MAST) at the Space Telescope Science Institute. The specific observations analyzed can be accessed via \dataset[doi: 10.17909/h3vj-zw90]{https://doi.org/10.17909/h3vj-zw90}.

The authors acknowledge the use of ChatGPT (https://chat.openai.com/, version 4.1) and doubao for editing the paper. 

\bibliographystyle{apj}
\bibliography{ms}

\clearpage

\appendix

\section{Synthetic $g$-band photometry}\label{appendix:synthetic}
Synthetic $g$-band photometry from spectra with the most reliable flux calibration in Section \ref{section:synthetic} is listed in Table \ref{tab:synthetic}.

\begin{table}[!h]
\centering
\caption{Synthetic $g$-band photometry}
\label{tab:synthetic}
\begin{tabular}{cccc}
\hline
Instr. & Date & Synthetic $g$-band (mag) & Error\\
\hline
LRIS & 2008-05-28 & 11.94 & 0.05 \\
DBSP & 2008-01-18 & 11.88 & 0.1  \\
DBSP & 2008-01-19 & 11.73 & 0.1  \\
DBSP & 2008-01-20 & 11.60 & 0.1  \\
DBSP & 2008-12-28 & 11.89 & 0.1  \\
DBSP & 2008-12-29 & 11.81 & 0.1  \\
DBSP & 2008-12-30 & 11.82 & 0.1  \\
XSH & 2010-04-07 & 12.03 & 0.1   \\
XSH & 2021-04-06 & 11.12 & 0.1   \\
XSH & 2021-04-08 & 11.53 & 0.1 \\
XSH & 2021-04-02 & 11.58 & 0.1 \\
HIRES & 2008-05-23 & 11.87 & 0.1 \\
STIS & 2000-05-07  &  11.65 & 0.01 \\
STIS & 2002-07-19 &  11.69 & 0.01 \\
STIS & 2010-01-28 &  11.56 & 0.01 \\
STIS & 2010-02-04 &  12.01 & 0.01 \\
STIS & 2010-05-28 &  11.81 & 0.01 \\
STIS & 2015-04-18 &  11.66 & 0.01 \\
\hline
\end{tabular}
\end{table}

\section{Additional veiling measurements}\label{appendix:veiling}
\citet{herczeg23} presented accretion rate measurements from 1167 high-resolution spectra.  In this paper, we add 31 new CHIRON spectra that were obtained in 2023, most during the TESS monitoring.  The veiling measurements from these spectra are presented in Table~\ref{tab:newveiling}.

\begin{table}
\centering
\caption{Veiling measurements from CHIRON observations in 2023}
\label{tab:newveiling}
\begin{tabular}{ccccc}
\hline
MJD & $r_{5050}$ & $\sigma(r)$ & $L_{\rm acc}$ & $M_{\rm acc}$ \\
& & & $(L_\odot)$ &  M$_\odot$ yr$^{-1}$\\
\hline
59960.26770 &      1.279 &      0.016 &     0.0799 &  4.454e-09 \\
59967.24484 &      0.571 &      0.009 &     0.0357 &  1.990e-09 \\
59975.29676 &      0.699 &      0.009 &     0.0437 &  2.434e-09 \\
60014.20552 &      0.336 &      0.020 &     0.0210 &  1.172e-09 \\
60015.16906 &      0.922 &      0.025 &     0.0576 &  3.211e-09 \\
60016.17687 &      0.827 &      0.013 &     0.0517 &  2.880e-09 \\
60017.14035 &      0.759 &      0.018 &     0.0474 &  2.643e-09 \\
60018.16355 &      0.557 &      0.012 &     0.0348 &  1.939e-09 \\
60019.11796 &      0.926 &      0.014 &     0.0578 &  3.224e-09 \\
60020.16611 &      0.759 &      0.017 &     0.0474 &  2.644e-09 \\
60021.16281 &      0.352 &      0.016 &     0.0220 &  1.228e-09 \\
60022.13396 &      0.335 &      0.012 &     0.0209 &  1.167e-09 \\
60023.15726 &      1.081 &      0.015 &     0.0676 &  3.765e-09 \\
60024.14446 &      0.590 &      0.018 &     0.0369 &  2.054e-09 \\
60025.21093 &      1.051 &      0.015 &     0.0657 &  3.660e-09 \\
60026.10258 &      0.871 &      0.018 &     0.0544 &  3.033e-09 \\
60027.10596 &      0.348 &      0.012 &     0.0218 &  1.213e-09 \\
60028.12458 &      0.792 &      0.068 &     0.0495 &  2.758e-09 \\
60029.13126 &      1.337 &      0.018 &     0.0835 &  4.655e-09 \\
60030.12944 &      1.040 &      0.023 &     0.0650 &  3.622e-09 \\
60031.15253 &      0.452 &      0.020 &     0.0282 &  1.573e-09 \\
60032.11392 &      0.840 &      0.030 &     0.0525 &  2.924e-09 \\
60033.12429 &      1.399 &      0.090 &     0.0874 &  4.871e-09 \\
60034.15969 &      0.932 &      0.083 &     0.0582 &  3.246e-09 \\
60035.08686 &      0.738 &      0.031 &     0.0461 &  2.570e-09 \\
60036.14270 &      0.529 &      0.012 &     0.0330 &  1.841e-09 \\
60037.07070 &      0.480 &      0.016 &     0.0300 &  1.671e-09 \\
60038.05662 &      0.331 &      0.012 &     0.0207 &  1.154e-09 \\
60039.11757 &      0.624 &      0.019 &     0.0390 &  2.172e-09 \\
60040.15208 &      0.411 &      0.016 &     0.0257 &  1.433e-09 \\
60041.11056 &      0.675 &      0.017 &     0.0422 &  2.352e-09 \\
\hline
\end{tabular}
\end{table}

\section{TESS 2021, 2023 and 2025 Accretion Time Series}\label{appendix:gaussian}
A detailed description of our burst fitting method is provided here. Each burst is fitted with the following Gaussian profile, with 
amplitude $A$, center $\mu$, and Gaussian width $\sigma$. We first fit the mass-accretion time series (after subtracting the constant baseline accretion rate) as a sum of individual major bursts. Initial parameter values for each major burst are chosen by visual inspection and allowed to vary during the fit using using the Python software lmfit \citep{matt24}. The number of major bursts is also treated as a free parameter. The preferred model is selected by matching the curve morphology and minimizing the residuals. Minor bursts are identified using a $2\times$RMS threshold and also modeled with a single Gaussian profile. Some minor bursts may be residuals from fits to major bursts.

\begin{figure*}[htbp]
	\centering
	\includegraphics[width=0.9\linewidth]{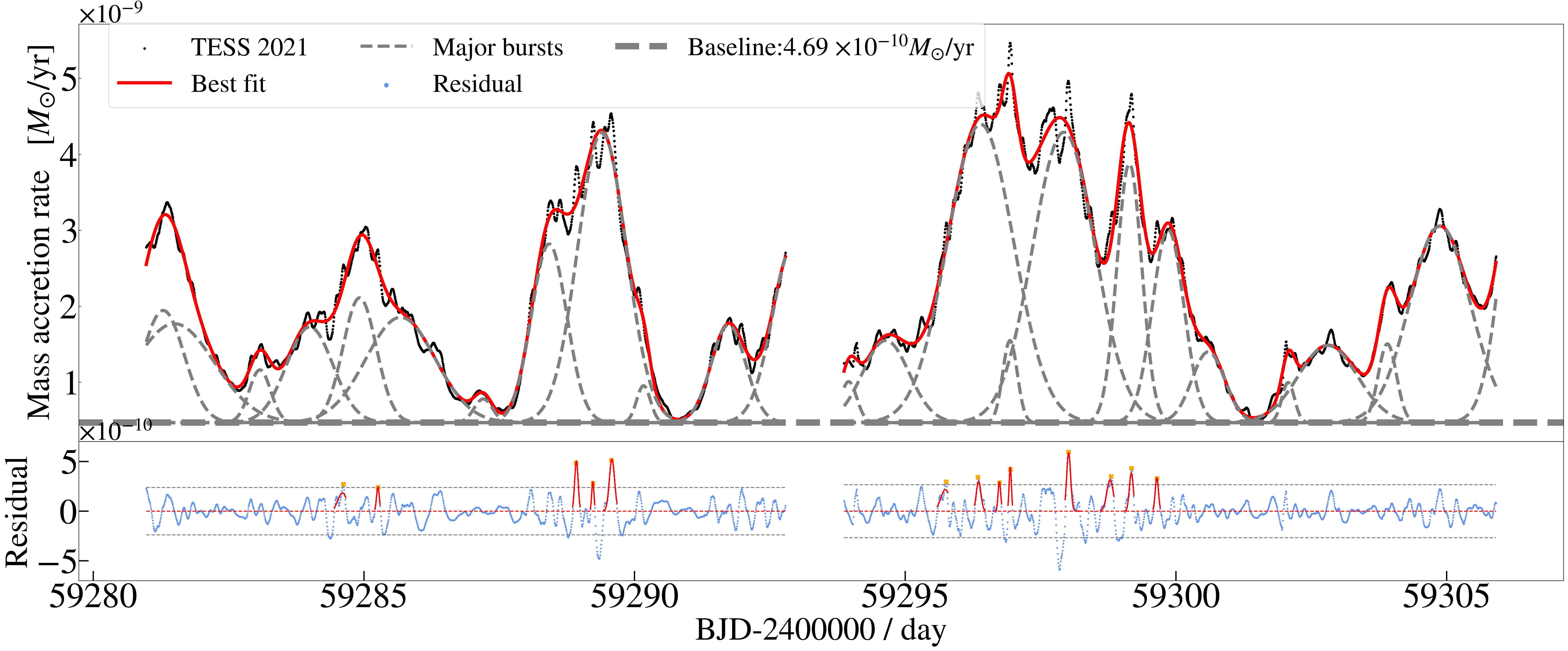}\label{fig:tess2021_fitting}
	\includegraphics[width=0.9\linewidth]{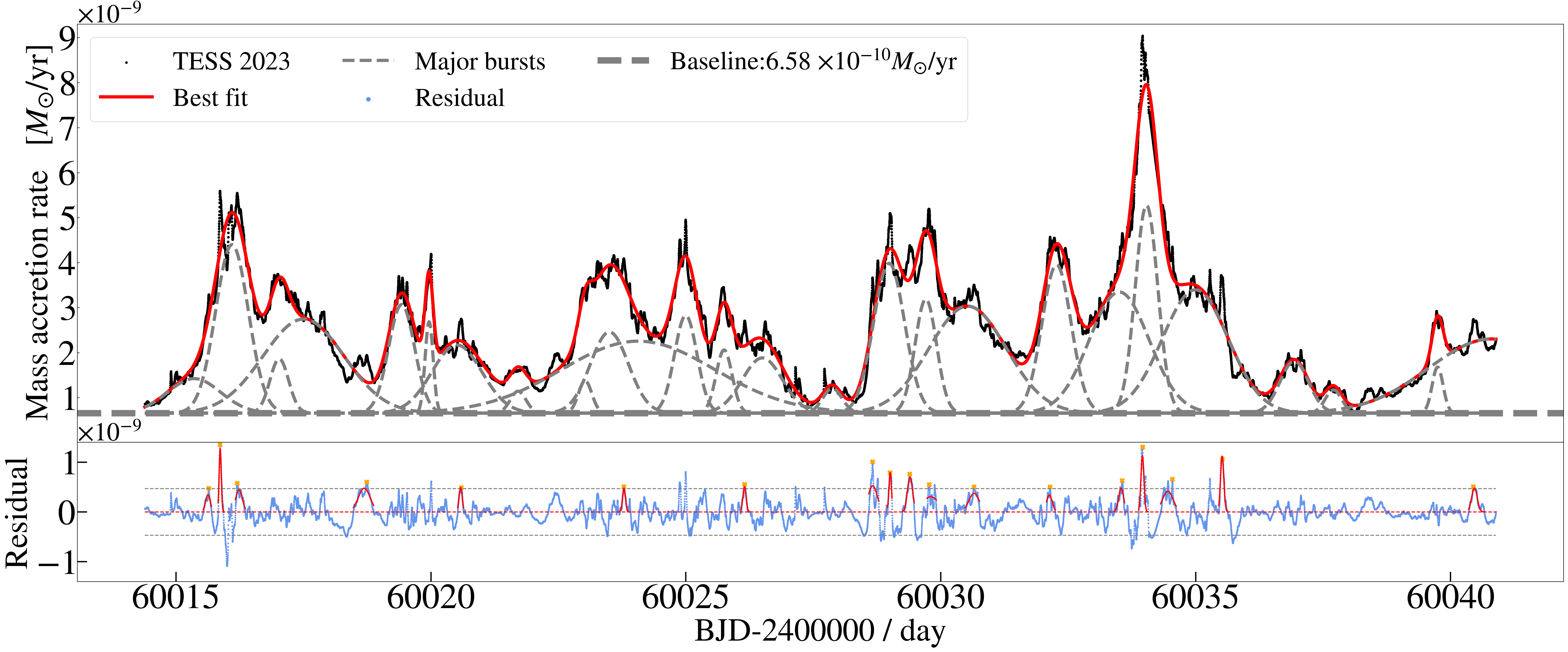}\label{fig:tess2023_fitting}
	\includegraphics[width=0.9\linewidth]{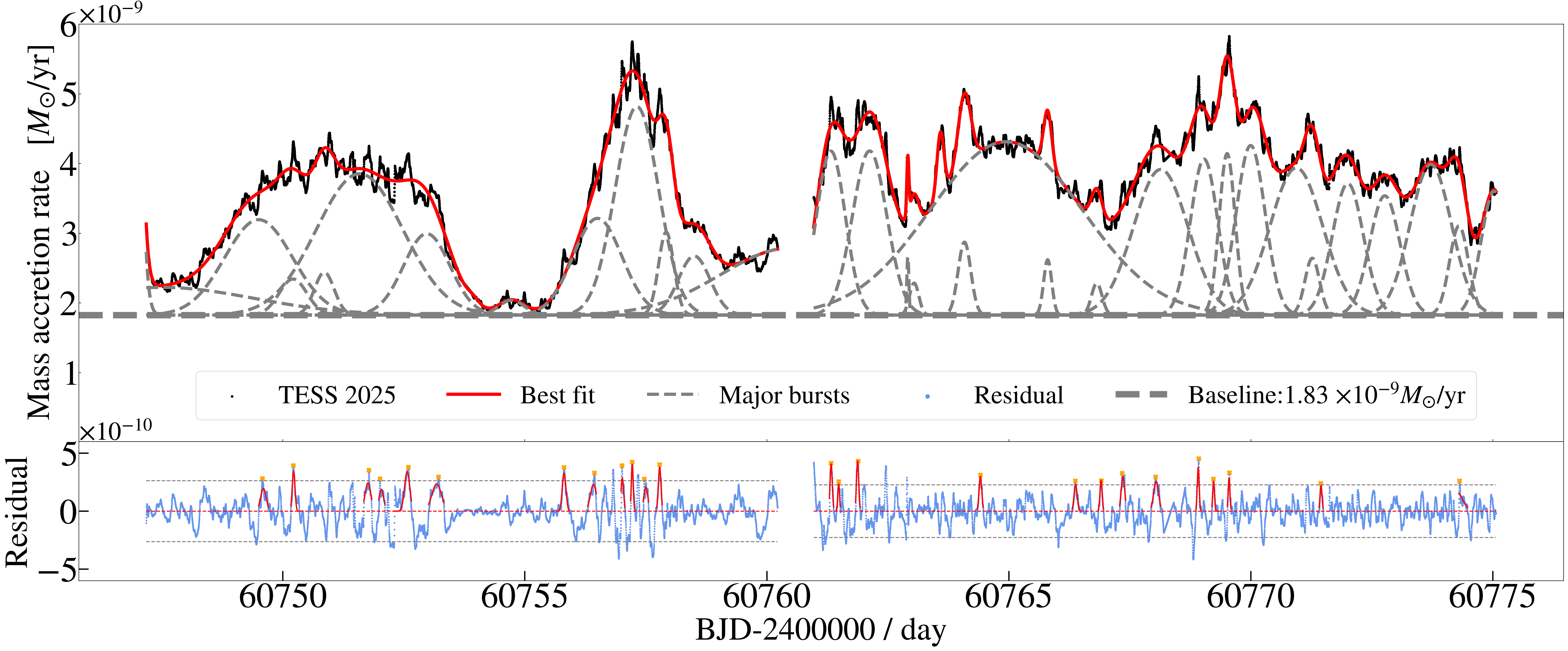}\label{fig:tess2025_fitting}
	\caption{The 2021, 2023, and 2025 TESS accretion time series, with points and lines following the presentation of the 2019 data in Figure ~\ref{fig:tess2019_fitting}.}
\end{figure*}

\clearpage

\section{TESS Structure Functions}\label{appendix:structure_function}

\begin{figure*}[htbp]
	\centering
	\includegraphics[width=.48\linewidth]{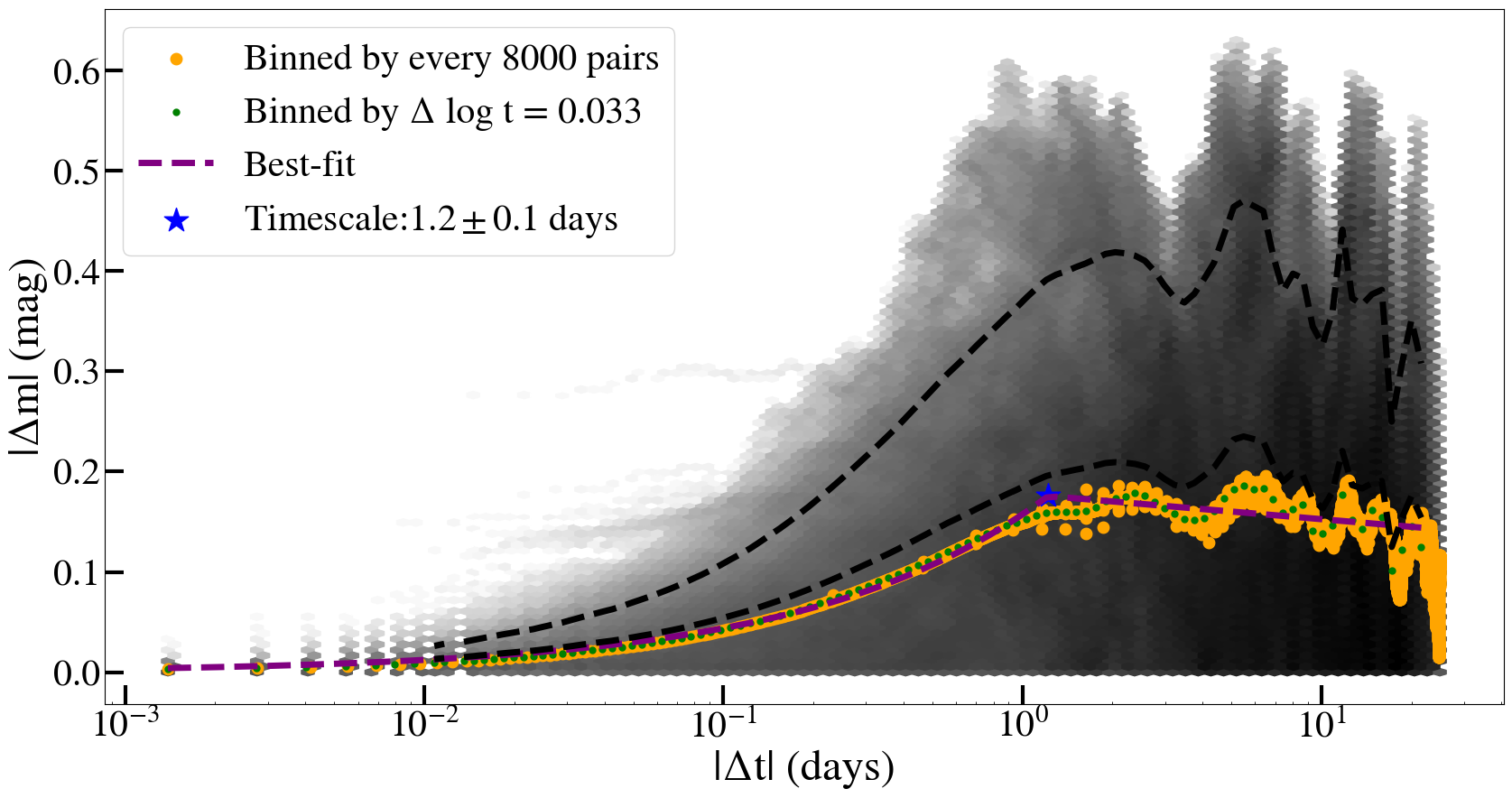}\hspace{5pt}
	\includegraphics[width=.48\linewidth]{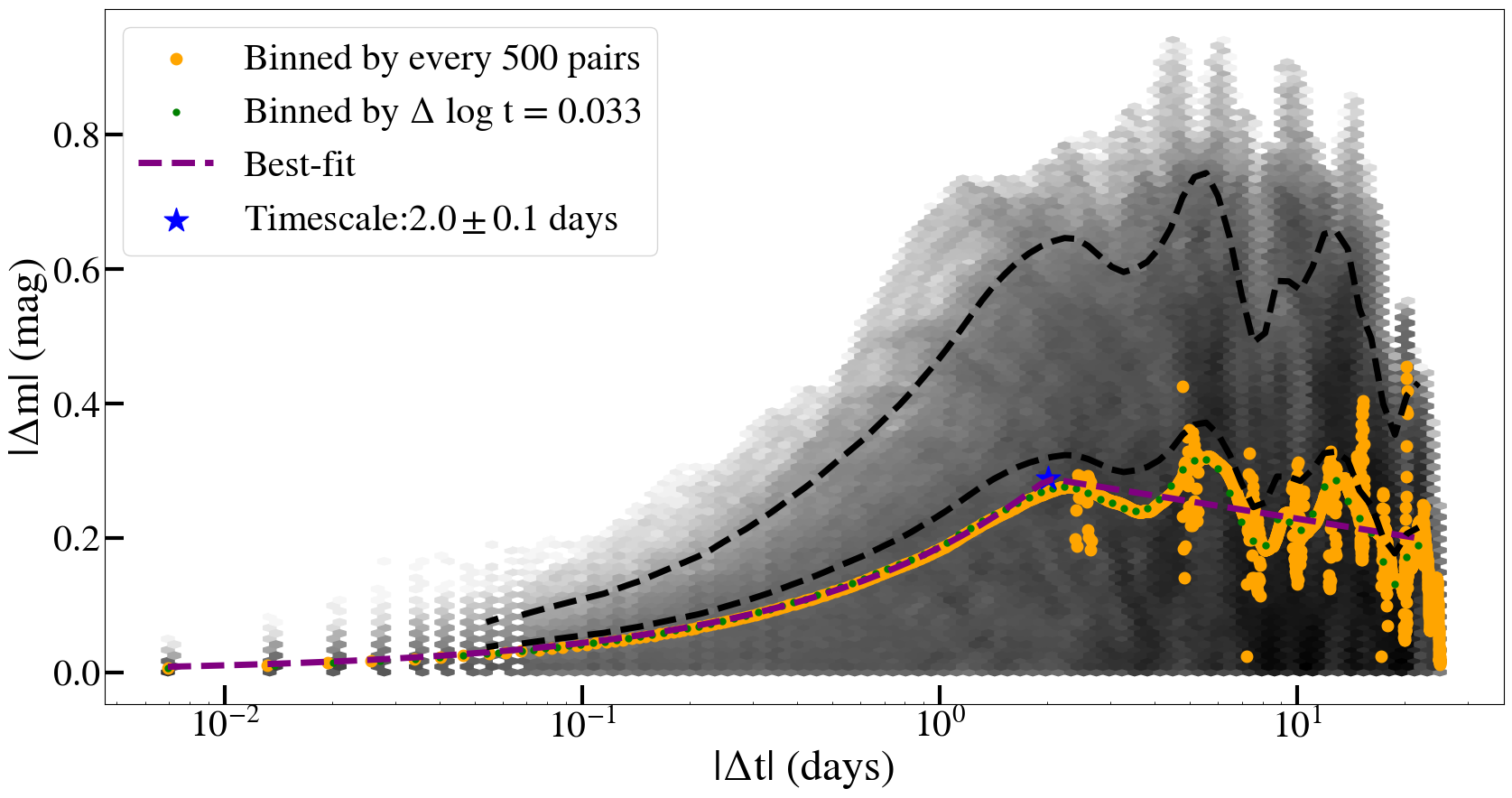}
	\includegraphics[width=.48\linewidth]{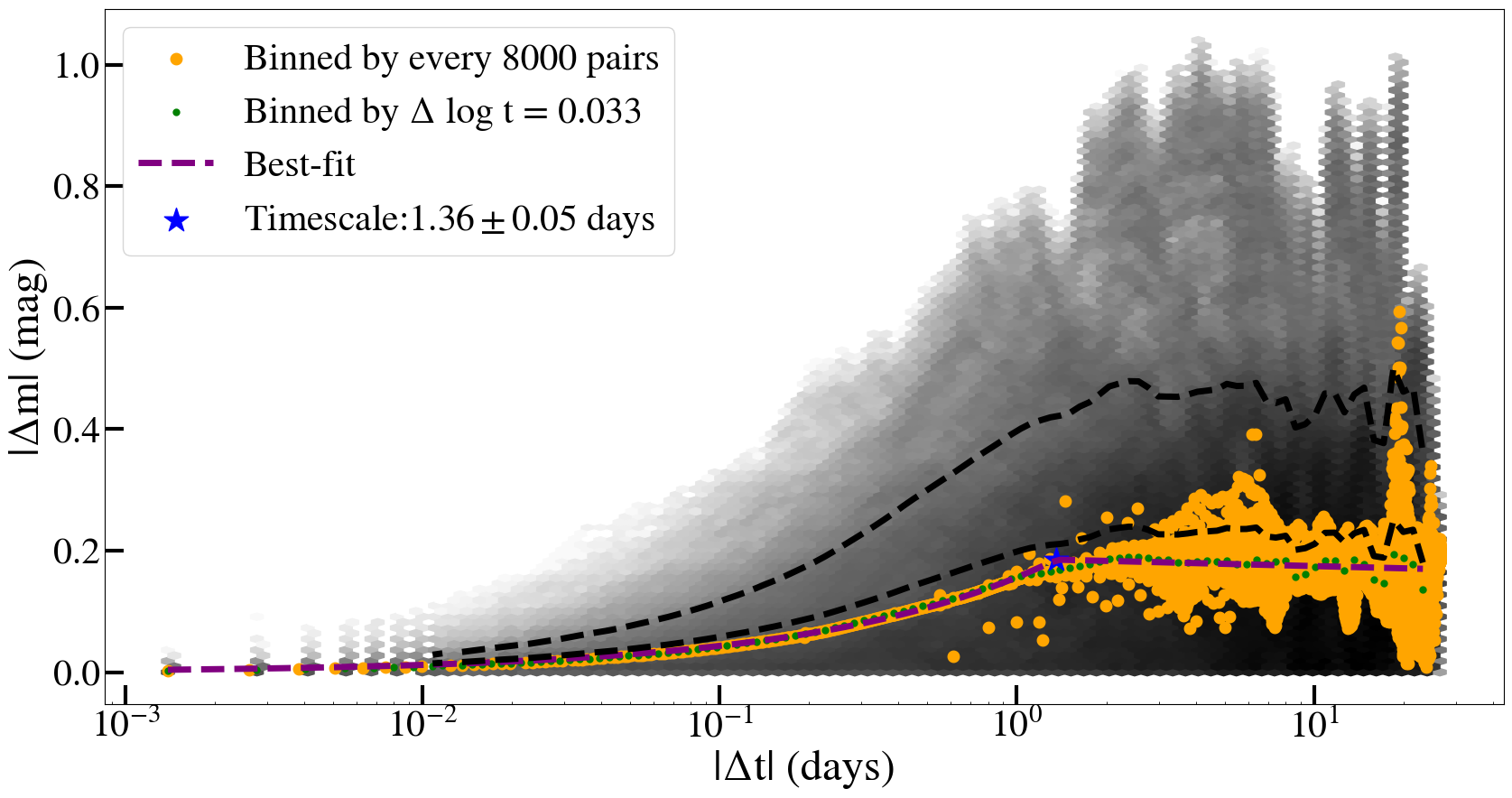}\hspace{5pt}
{\includegraphics[width=.48\linewidth]{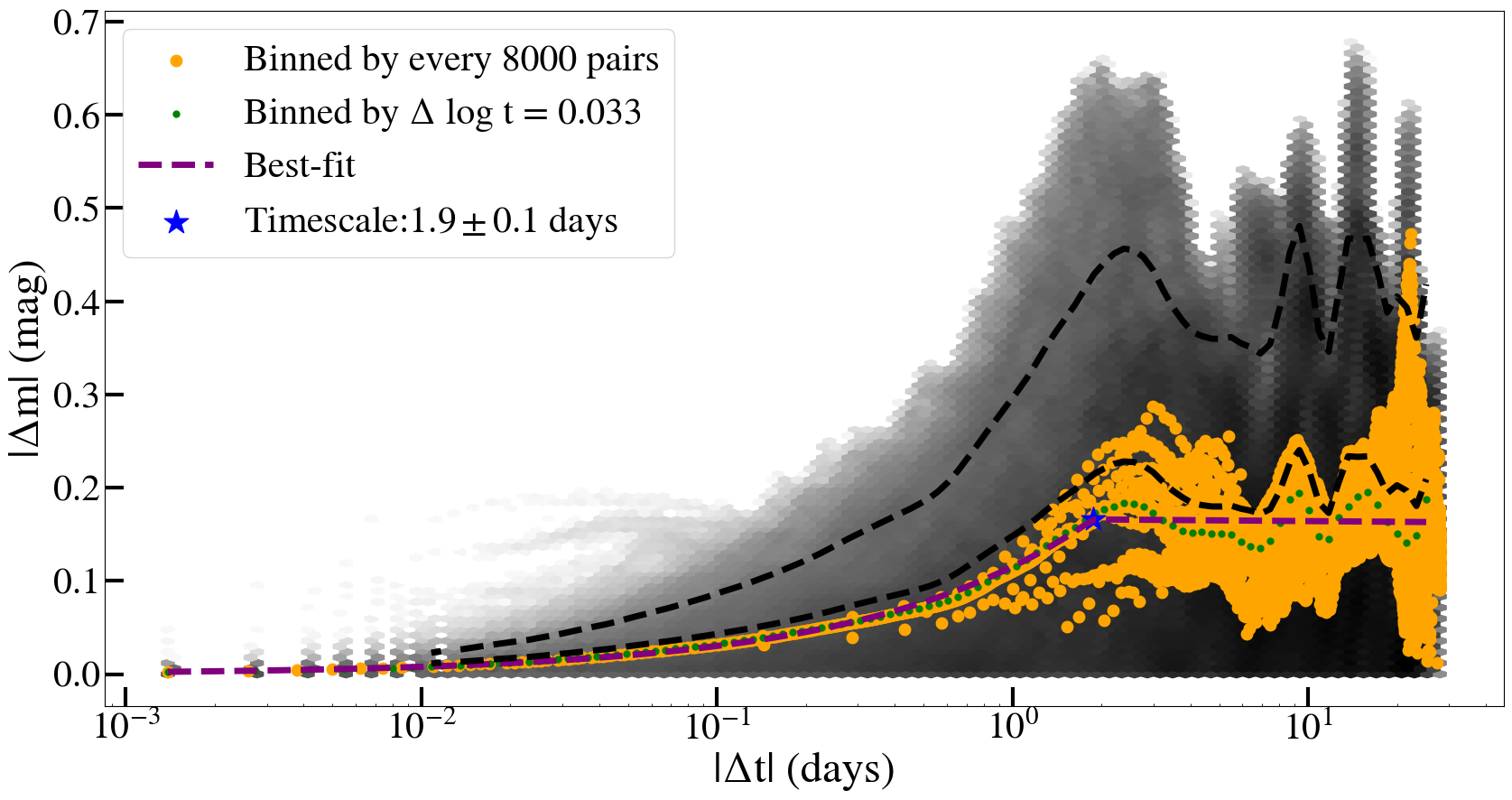}}
	\caption{TESS structure functions and fit results, with data points and lines following the structure function of ASAS-SN data presented in Figure~\ref{fig:structure_function_asassn}.}    
    \label{fig:structureTESS}
\end{figure*}

The structure function for the four TESS light curves are presented in Figure~\ref{fig:structureTESS}.  The best-fit reset timescales range from 1.2--2.0 days.

\section{Long-term Variability}
\label{appendix:long-term}

We divided the ASAS-SN monitoring into 30-day periods and calculated the average magnitude and corresponding accretion rate for each period with more than 15 data points.  The results from this analysis are presented in Table~\ref{tab:averages}.

\begin{longtable}{ccccc}
\caption{Monthly averages of ASAS-SN monitoring}
\label{tab:averages}
\\
\hline
\multicolumn{2}{c} {Date Interval}& avg $g$ & $\sigma(g)$ & avg $\dot{M}_{acc} \ [10^{-9} $M$_{\odot}$~yr$^{-1}$]\\ \hline
\endfirsthead

\multicolumn{5}{c}
{{\bfseries \tablename\ \thetable{} -- Continued from previous page}} \\
\hline 
\multicolumn{2}{c} {Date Interval}& avg $g$ & $\sigma(g)$ & avg $\dot{M}_{acc} \ [10^{-9} $M$_{\odot}$~yr$^{-1}$]\\
\hline
\endhead

\hline \multicolumn{5}{c}{{Continued on next page}} \\ \hline
\endfoot

\hline \hline
\endlastfoot

\hline
2017-12-01 & 2017-12-30 & 11.58 & 0.26 & 2.67 \\
2018-01-03 & 2018-01-28 & 11.49 & 0.20 & 3.05 \\
2018-01-29 & 2018-03-01 & 11.50 & 0.22 & 3.01 \\
2018-03-02 & 2018-03-31 & 11.56 & 0.19 & 2.75 \\
2018-04-01 & 2018-04-30 & 11.49 & 0.24 & 3.05 \\
2018-04-30 & 2018-05-28 & 11.37 & 0.19 & 3.55 \\
2018-05-31 & 2018-06-23 & 11.50 & 0.25 & 3.01 \\
2018-10-28 & 2018-11-25 & 11.49 & 0.19 & 3.05 \\
2018-11-27 & 2018-12-25 & 11.39 & 0.23 & 3.47 \\
2018-12-27 & 2019-01-25 & 11.42 & 0.18 & 3.34 \\
2019-01-25 & 2019-02-24 & 11.36 & 0.20 & 3.60 \\
2019-02-25 & 2019-03-26 & 11.48 & 0.19 & 3.09 \\
2019-03-27 & 2019-04-24 & 11.53 & 0.17 & 2.88 \\
2019-04-25 & 2019-05-24 & 11.55 & 0.17 & 2.79 \\
2019-05-31 & 2019-06-24 & 11.47 & 0.20 & 3.13 \\
2019-06-26 & 2019-07-22 & 11.44 & 0.18 & 3.26 \\
2019-11-22 & 2019-12-20 & 11.47 & 0.14 & 3.13 \\
2019-12-22 & 2020-01-20 & 11.42 & 0.17 & 3.34 \\
2020-01-21 & 2020-02-19 & 11.36 & 0.20 & 3.60 \\
2020-02-20 & 2020-03-17 & 11.46 & 0.20 & 3.17 \\
2020-10-26 & 2020-11-15 & 11.67 & 0.13 & 2.29 \\
2020-11-18 & 2020-12-15 & 11.68 & 0.15 & 2.24 \\
2020-12-16 & 2021-01-12 & 11.60 & 0.17 & 2.58 \\
2021-01-14 & 2021-02-13 & 11.70 & 0.23 & 2.16 \\
2021-02-14 & 2021-03-14 & 11.72 & 0.19 & 2.08 \\
2021-03-17 & 2021-04-14 & 11.56 & 0.29 & 2.75 \\
2021-04-15 & 2021-05-14 & 11.59 & 0.26 & 2.62 \\
2021-05-16 & 2021-06-12 & 11.65 & 0.21 & 2.37 \\
2021-06-15 & 2021-07-12 & 11.67 & 0.27 & 2.29 \\
2021-07-14 & 2021-08-14 & 11.52 & 0.29 & 2.92 \\
2021-10-14 & 2021-11-09 & 11.20 & 0.22 & 4.27 \\
2021-11-11 & 2021-12-09 & 11.51 & 0.22 & 2.96 \\
2021-12-14 & 2022-01-09 & 11.21 & 0.31 & 4.23 \\
2022-01-12 & 2022-02-07 & 11.33 & 0.21 & 3.72 \\
2022-02-08 & 2022-03-10 & 11.26 & 0.27 & 4.02 \\
2022-03-11 & 2022-04-09 & 11.40 & 0.20 & 3.43 \\
2022-04-12 & 2022-05-08 & 11.27 & 0.18 & 3.98 \\
2022-05-10 & 2022-06-07 & 11.32 & 0.18 & 3.77 \\
2022-06-09 & 2022-07-06 & 11.46 & 0.24 & 3.17 \\
2022-07-08 & 2022-08-04 & 11.47 & 0.18 & 3.13 \\
2022-10-18 & 2022-11-04 & 11.52 & 0.19 & 2.92 \\
2022-11-05 & 2022-12-05 & 11.48 & 0.19 & 3.09 \\
2022-12-07 & 2023-01-04 & 11.54 & 0.18 & 2.84 \\
2023-01-06 & 2023-02-02 & 11.64 & 0.23 & 2.41 \\
2023-02-04 & 2023-03-04 & 11.61 & 0.16 & 2.54 \\
2023-03-06 & 2023-04-04 & 11.51 & 0.27 & 2.96 \\
2023-04-04 & 2023-05-04 & 11.62 & 0.29 & 2.50 \\
2023-05-06 & 2023-06-03 & 11.54 & 0.31 & 2.84 \\
2023-06-05 & 2023-07-01 & 11.50 & 0.29 & 3.01 \\
2023-07-03 & 2023-08-01 & 11.65 & 0.31 & 2.37 \\
2023-10-31 & 2023-11-30 & 11.57 & 0.26 & 2.71 \\
2023-12-08 & 2023-12-14 & 11.56 & 0.15 & 2.75 \\
2024-01-03 & 2024-01-28 & 11.71 & 0.18 & 2.12 \\
2024-01-30 & 2024-02-27 & 11.70 & 0.18 & 2.16 \\
2024-02-29 & 2024-03-29 & 11.70 & 0.14 & 2.16\\
2024-03-30 & 2024-04-27 & 11.76 & 0.20 & 1.91 \\
2024-04-28 & 2024-05-28 & 11.58 & 0.20 & 2.67 \\
2024-05-29 & 2024-06-27 & 11.63 & 0.19 & 2.46 \\
2024-06-28 & 2024-07-25 & 11.54 & 0.21 & 2.84 \\
2024-11-03 & 2024-11-24 & 11.45 & 0.18 & 3.22 \\
2024-11-25 & 2024-12-22 & 11.60 & 0.20 & 2.58 \\
2024-12-25 & 2025-01-22 & 11.49 & 0.18 & 3.05 \\
2025-01-23 & 2025-02-22 & 11.54 & 0.21 & 2.84 \\
2025-02-23 & 2025-03-23 & 11.55 & 0.24 & 2.79 \\
2025-03-24 & 2025-04-23 & 11.38 & 0.12 & 3.52\\
2025-04-24 & 2025-05-17 & 11.43 & 0.16 & 3.31\\
\hline
\end{longtable}

\end{CJK*}

\end{document}